\documentclass[reprint,superscriptaddress,amsmath,amssymb,aps,prd]{revtex4-2}

\usepackage{graphicx}
\usepackage{dcolumn}
\usepackage{bm}
\usepackage[colorlinks=true,citecolor=blue,urlcolor=blue,linkcolor=blue]{hyperref}
\usepackage{orcidlink}
\usepackage{booktabs}
\usepackage{multirow}
\usepackage{tikz}
\usepackage{xcolor}
\usetikzlibrary{arrows.meta,calc,backgrounds,fit,positioning}
\usepackage{aas_macros}
\usepackage{subcaption}

\newif\ifDraft
\Drafttrue
\definecolor{DarkGreen}{rgb}{0.0,0.45,0}
\definecolor{Sakujo}{rgb}{0.9,0.5,0.9}
\definecolor{Grey}{rgb}{0.7,0.7,0.7}
\ifDraft

\else

\fi

\begin{document}
\title{Deep Learning Recognition of Gravitational Waves \\
from Three-Dimensional Core-Collapse Supernova Simulations: \\
Fourier and Hilbert–Huang Representations}

\author{Yusuke Sakai\orcidlink{0000-0001-8810-4813}}
\affiliation{Research Center for Space Science, Advanced Research Laboratories, Tokyo City University, 1-28-1 Tamazutsumi, Setagaya-ku, Tokyo 158-8557, Japan}
\affiliation{Graduate School of Information and Data Science and Department of Design and Data Science, Tokyo City University, 3-3-1 Ushikubo-Nishi, Tsuzuki-ku, Yokohama, Kanagawa 224-8551, Japan}
\email{y.sakai@tcu.ac.jp}

\author{Takahiro S. Yamamoto\orcidlink{0000-0002-8181-924X}}
\affiliation{Research Center for the Early Universe, The University of Tokyo, 7-3-1 Hongo, Bunkyo-ku, Tokyo 113-0033, Japan}

\author{Christopher All\'en\'e\orcidlink{0009-0001-3859-5420}}
\affiliation{Research Center for Space Science, Advanced Research Laboratories, Tokyo City University, 1-28-1 Tamazutsumi, Setagaya-ku, Tokyo 158-8557, Japan}

\author{Kazuki Sakai\orcidlink{0009-0000-2457-3901}}
\affiliation{Department of Electronic Control Engineering, National Institute of Technology, Nagaoka College, 888 Nishikatakai, Nagaoka, Niigata 940-8532, Japan}

\author{Ken-ichi Oohara\orcidlink{0000-0002-7518-6677}}
\affiliation{Nambu Yoichiro Institute of Theoretical and Experimental Physics (NITEP),
Osaka Metropolitan University, 3-3-138 Sugimoto-cho, Sumiyoshi-ku, Osaka City,
Osaka 558-8585, Japan}
\affiliation{Graduate School of Science and Technology, Niigata University, 8050 Ikarashi-2-no-cho, Nishi-ku, Niigata 950-2181, Japan}

\author{Yoshikazu Terada\orcidlink{0000-0002-4509-1108}}
\affiliation{Graduate School of Engineering Science, The University of Osaka, 1-3 Machikaneyama, Toyonaka, Osaka 560-8531, Japan}
\affiliation{Center for Advanced Intelligence Project, RIKEN, 2-1 Hirosawa, Wako, Saitama 351-0198, Japan}

\author{Marco Meyer-Conde\orcidlink{0000-0003-2230-6310}}
\affiliation{Research Center for Space Science, Advanced Research Laboratories, Tokyo City University, 1-28-1 Tamazutsumi, Setagaya-ku, Tokyo 158-8557, Japan}

\author{Hirotaka Takahashi\orcidlink{0000-0003-0596-4397}}
\affiliation{Research Center for Space Science, Advanced Research Laboratories, Tokyo City University, 1-28-1 Tamazutsumi, Setagaya-ku, Tokyo 158-8557, Japan}
\affiliation{Graduate School of Information and Data Science and Department of Design and Data Science, Tokyo City University, 3-3-1 Ushikubo-Nishi, Tsuzuki-ku, Yokohama, Kanagawa 224-8551, Japan}
\affiliation{Earthquake Research Institute, The University of Tokyo, 1-1-1 Yayoi, Bunkyo-ku, Tokyo 113-0032, Japan}
\email{hirotaka@tcu.ac.jp}

\begin{abstract}
Gravitational-wave (GW) signals from Galactic core-collapse supernovae (CCSNe) provide a unique probe of the explosion dynamics. However, their stochastic and diverse waveform morphology makes astrophysical interpretation challenging. 
As a prerequisite for extracting physical information from future CCSN observations, it is important to establish whether simulated waveforms corresponding to different explosion scenarios can be reliably distinguished.
%As a first step toward understanding the physics of CCSNe, it is essential to determine whether waveforms obtained from numerical simulations can be reliably distinguished. 
We study a catalog of 51 waveforms from three-dimensional CCSN simulations, projected onto the three interferometers of the Einstein Telescope. The signals are processed using the coherent WaveBurst (cWB) pipeline, a widely used burst-search algorithm, to reconstruct candidate GW signals. We also apply two complementary time--frequency analysis methods as preprocessing: the short-time Fourier transform (STFT) and the Hilbert--Huang transform (HHT). We use ConvNeXt V2 and an axial attention transformer to classify the STFT and HHT representations, respectively. Conditional on the availability of a cWB reconstruction, the STFT and HHT classifiers achieve accuracies of 98.8\% and 95.2\%, respectively, at 100~kpc for the 19 of 51 waveform classes represented in the test sample. Selecting predictions with high confidence yields accuracies  approximately 100\% at every evaluated distance. These results demonstrate that deep learning can help identify the underlying CCSN waveform model for Galactic CCSNe detectable by future third-generation gravitational-wave detectors.
\end{abstract}

\maketitle

\section{Introduction}
Since the beginning of gravitational-wave astronomy with the LIGO~\cite{ligo}, Virgo~\cite{virgo}, and KAGRA~\cite{kagra} (LVK) collaboration, 390 compact binary coalescence candidates have been reported to date~\cite{gwtc-1, gwtc-2, gwtc-2.1, gwtc-3, gwtc-4, gwtc-5, gwtc-5.1}.
The binary neutron star (BNS) merger event GW170817~\cite{PhysRevLett.119.161101}, followed by the probable BNS event GW190425~\cite{2020ApJ...892L...3A}, has demonstrated the scientific potential of gravitational-wave astronomy. In particular, the multi-messenger observation of GW170817 provided new insights into neutron-star physics and r-process nucleosynthesis~\cite{2017ApJ...848L..12A}, while post-merger gravitational waves contributed to further constrain the neutron-star equation of state (EOS)~\cite{2018PhRvL.121p1101A}.
In contrast, gravitational waves from core-collapse supernovae (CCSNe) have not yet been detected, although their future observation is expected to provide unique insights into the explosion mechanism and the properties of proto-neutron stars (PNSs).
CCSNe are the explosions of massive stars and represent uniquely information-rich multi-messenger events, in which GWs, neutrinos, and electromagnetic signals provide complementary information on the explosion mechanism~\cite{2016MNRAS.461.3296N}.
Beyond the LVK detector network, next-generation gravitational-wave telescopes are currently under development. The Einstein Telescope (ET)~\cite{2010CQGra..27s4002P, 2026JCAP...03..081A}, a 10-km underground detector in Europe, and the Cosmic Explorer (CE)~\cite{2019BAAS...51g..35R, 2021arXiv210909882E}, a 40-km ground-based detector in the United States, are designed to achieve substantially higher sensitivities.

Although the explosion mechanism of CCSNe remains incompletely understood~\cite{2012ARNPS..62..407J}, gravitational waves (GWs), which are emitted directly by nonspherical bulk motions in the stellar core, provide a unique probe of the explosion dynamics that complements neutrino and electromagnetic observations.
To date, multi-dimensional numerical simulations have successfully produced explosions under a variety of conditions~\cite{szczepanczyk2021detecting}.
These studies also suggest that the current gravitational-wave detector network may detect CCSN GWs from Galactic events within a distance of approximately 10~kpc.
In contrast, ET is expected to substantially extend the detection range. For a representative $15\,M_{\odot}$ progenitor, the detection horizon is estimated to reach $\sim100$~kpc, providing essentially complete coverage of the Milky Way and partial coverage of the Magellanic Clouds~\cite{ccsn-et}.
Such an extended observational reach will provide opportunities not only for detecting CCSN GWs but also for investigating the physical information encoded in their waveform features.
The physical conditions and explosion dynamics, including the progenitor mass, rotation, magnetic fields, standing accretion shock instability (SASI)~\cite{Roberts2012, Kuroda2016}, and convection, differ substantially among models. Consequently, the resulting GW signals exhibit a wide diversity of waveform morphologies and time--frequency characteristics.
Nevertheless, the PNS exhibits characteristic oscillation eigenmodes, such as fundamental and gravity modes, whose frequencies are connected to the physical parameters of the PNS. These universal relations~\cite{Torres2021} enable the underlying PNS properties to be inferred from GW observations. Therefore, extracting features from GW signals provides important clues to understanding the explosion mechanism of CCSNe.

Unlike compact binary coalescence signals, CCSN GW signals are stochastic and highly complex, making it challenging to construct sufficiently accurate waveform template families for conventional matched-filtering techniques.
Consequently, time--frequency analysis, represented by the short-time Fourier transform (STFT) and wavelet transform, is a standard approach for investigating these unmodeled signals.
GW burst-search pipelines, such as BayesWave~\cite{2021PhRvD.103d4006C}, oLIB~\cite{2017PhRvD..95j4046L}, PySTAMPAS~\cite{2021arXiv210810033M}, MLy~\cite{mly}, and coherent WaveBurst (cWB)~\cite{Klimenko2008,Klimenko2016,Drago2021,2022APS..APRE17004K}, search for GW transients without requiring a specific waveform morphology.
Notably, cWB was one of the pipelines that identified the first gravitational-wave event, GW150914, from a binary black hole merger in a low-latency search~\cite{PhysRevLett.116.061102}.
Time--frequency analysis has been widely used to investigate the physical characteristics of CCSN GWs, including SASI and PNS oscillations~\cite{2023ApJ...959...21P}. Furthermore, advanced analysis techniques have been proposed to improve signal characterization~\cite{2024MNRAS.529.3235Y}. In particular, the Hilbert--Huang transform (HHT) has been shown to be effective for extracting SASI signatures~\cite{Takeda2021}.

Recently, deep learning has been applied to CCSN GW analysis, demonstrating that CCSN waveforms can be recognized with high accuracy~\cite{2023PhRvD.108l3033S, PhysRevD.109.063019, sun2026contrastive, Casallas-Lagos:2026met}.
Previous studies have primarily demonstrated the classification using subsets of CCSN waveforms.
The performance of these approaches on larger, more diverse catalogs requires further investigation.
In this study, we investigate whether neural networks with a large number of trainable parameters can distinguish among a large set of CCSN waveforms spanning diverse physical configurations.
We further examine whether the HHT can also provide an effective representation for deep-learning-based classification. Using two distinct analysis pipelines based on different preprocessing procedures, time--frequency representations, and neural network architectures, we investigate whether the STFT- and HHT-based approaches yield consistent classification results.

The remainder of this paper is organized as follows. Section~\ref{sec2} describes the numerical waveform catalog, detector configuration, and signal preprocessing. Section~\ref{sec3} introduces the time--frequency representations based on the STFT and HHT. Section~\ref{sec4} presents the deep-learning architectures, dataset construction, and training procedure. Section~\ref{sec5} presents the classification results, including the performance with and without cWB reconstruction. Section~\ref{sec6} discusses the classification performance and the physical interpretation of the learned features. Finally, Sec.~\ref{sec7} summarizes the main conclusions of this study.

%Future concepts have also been proposed to explore the low-frequency gravitational-wave band (below a few hertz), which is challenging to observe with current ground-based detectors. These include atom interferometers, superconducting gravity gradiometers, and the torsion-bar antenna (TOBA). In addition, space-based gravitational-wave observatories, including the Laser Interferometer Space Antenna (LISA)~\cite{...} and the Deci-hertz Interferometer Gravitational-wave Observatory (DECIGO)~\cite{...}, are planned to extend gravitational-wave observations to the millihertz--decihertz frequency band.

%3. detection of CCSNe will be achieved by multi-band EM observation and neutrino. timing of the neutrino would be close to the emission of GWs if SASI exsists. So the time of occurrence of GWs can be determined from other observations. First, it is important to estimate the type of CCSNe from GW solely. Furthermore, showing consistent results in the analysis of GW, electromagnetic waves, and neutrinos will lead to a better understanding of physics.\\\\

\section{Data and signal construction}\label{sec2}
\begin{figure}[t]
\centering
\begin{tikzpicture}[
>=Stealth,
box/.style={
    draw,
    rounded corners=2mm,
    minimum width=3cm,
    minimum height=7mm,
    align=center,
    font=\small
}]
% Main flow
\node[box] (wave)
{CCSN GWs\\$(h_{+},\,h_{\times})$};
\node[box, below=5mm of wave] (et)
{Einstein Telescope};
\node[box, below=5mm of et] (bp)
{Band-pass filter\\100--1200 Hz};
\node[box, below=5mm of bp] (cwb)
{(Optional)\\cWB};

% Branch
\coordinate (split) at ($(cwb.south)+(0,-5mm)$);

\node[box, below left=6mm and 7mm of split] (stft)
{STFT};
\node[box, below right=6mm and 7mm of split] (hht)
{HHT};
\node[box, below=10mm of stft] (conv)
{ConvNeXt V2};
\node[box, below=10mm of hht] (trans)
{Transformer};

% Final node
\node[
box,
minimum width=5.5cm,
below=20mm of $(conv)!0.5!(trans)$
] (cls)
{Waveform classification};

% Arrows
\draw[->] (wave) -- (et);
\draw[->] (et) -- (bp);
\draw[->] (bp) -- (cwb);
% split
\draw[->] (cwb) -- (split);
\draw (split) -- ($(stft.north)+(0,6mm)$);
\draw (split) -- ($(hht.north)+(0,6mm)$);
\draw[->] ($(stft.north)+(0,6mm)$) -- (stft);
\draw[->] ($(hht.north)+(0,6mm)$) -- (hht);
\draw[->] (stft) -- (conv);
\draw[->] (hht) -- (trans);
% separate arrows
\draw[->] (conv.south) -- ++(0,-5mm) -| ([xshift=-5mm]cls.north);
\draw[->] (trans.south) -- ++(0,-5mm) -| ([xshift=5mm]cls.north);
\end{tikzpicture}
\caption{
Workflow of the proposed CCSN waveform classification framework.
The waveforms are band-pass filtered to the characteristic frequency range of CCSNe and, optionally, reconstructed from the three ET detectors using cWB. The resulting signals are then transformed into STFT or HHT time--frequency representations for waveform classification.
}
\label{fig:workflow}
\end{figure}
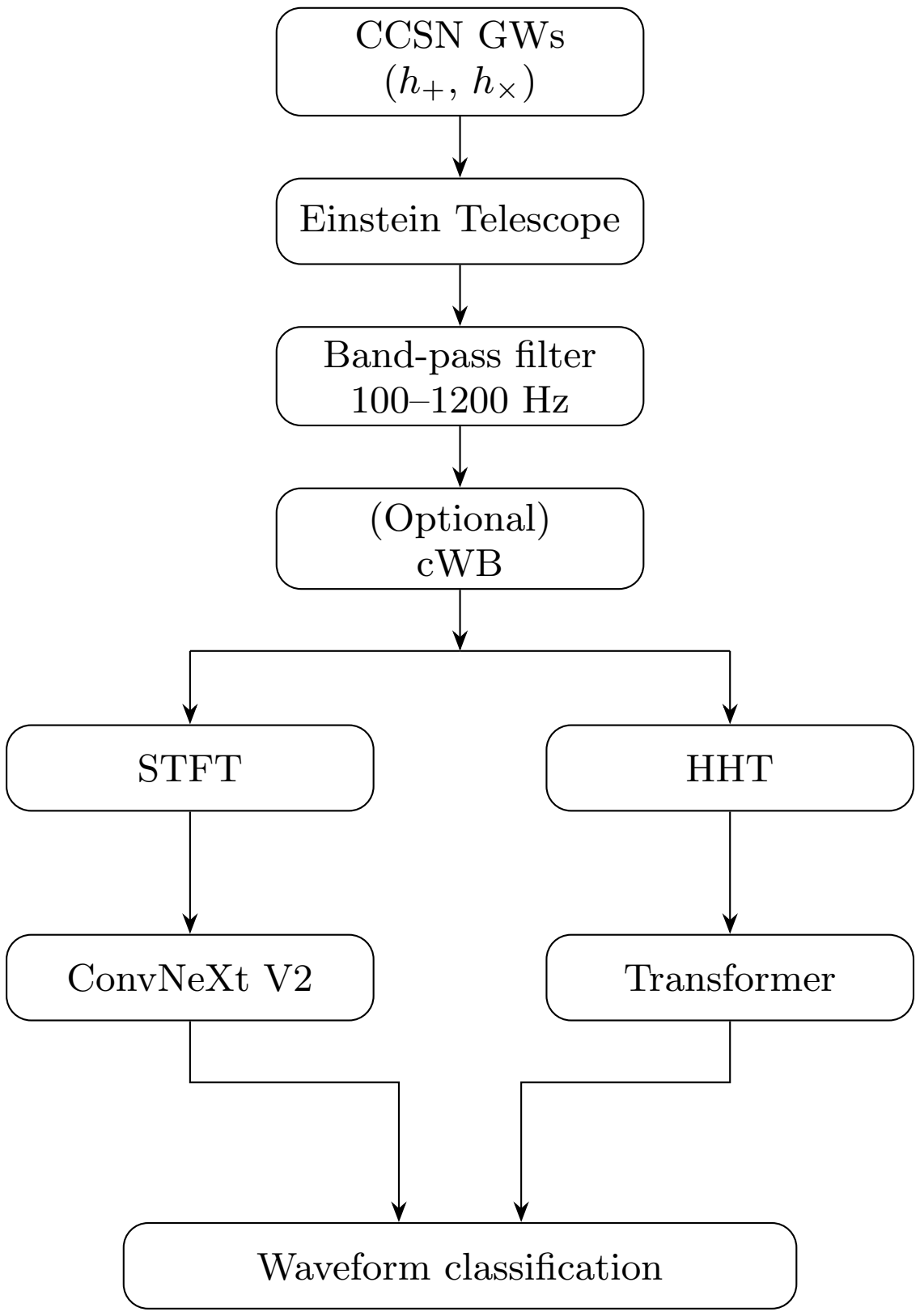

Figure~\ref{fig:workflow} illustrates the overall workflow of the waveform classification framework. This section describes the waveform catalog, detector configuration, and signal preprocessing used in this study.

\subsection{Numerical waveform catalog}
As a brief overview, CCSNe occur when massive stars with zero-age main-sequence (ZAMS) masses typically exceeding $8\,M_\odot$ undergo gravitational core collapse at the end of their evolution, leading to the formation of a proto-neutron star (PNS). During the post-bounce phase, the PNS undergoes dynamical oscillations on millisecond timescales. The resulting gravitational-wave emission is characterized by a low-frequency component associated with standing accretion shock instability (SASI)~\cite{Roberts2012, Kuroda2016} and convection, together with a dominant high-frequency component produced by PNS surface g-mode oscillations. As the PNS contracts due to continued accretion, the characteristic g-mode frequency increases with time, reaching frequencies of up to approximately $1\,\mathrm{kHz}$ during the first second after core bounce~\cite{Torres2021}.

\citeauthor{szczepanczyk2021detecting}~\cite{szczepanczyk2021detecting} provide a catalog of CCSN waveforms obtained from both two-dimensional and three-dimensional (3D) numerical simulations. We use only waveforms from fully 3D simulations, which provide both the plus and cross polarizations and therefore more realistically represent GW signals in actual observations.
We extend this catalog by including recently published 3D CCSN simulation waveforms, as summarized in Table~\ref{tab:waveforms}. In total, 51 waveforms from multiple simulation studies are used.
Each waveform is represented by the two polarizations, $h_+(t; \theta, \phi)$ and $h_\times(t; \theta, \phi)$, as functions of the source orientation $(\theta, \phi)$.  The waveforms are stored at a sampling frequency of 16,384~Hz and resampled to 4096~Hz for analysis.

\begin{figure}[h!]
    \centering
    \includegraphics[width=\columnwidth]{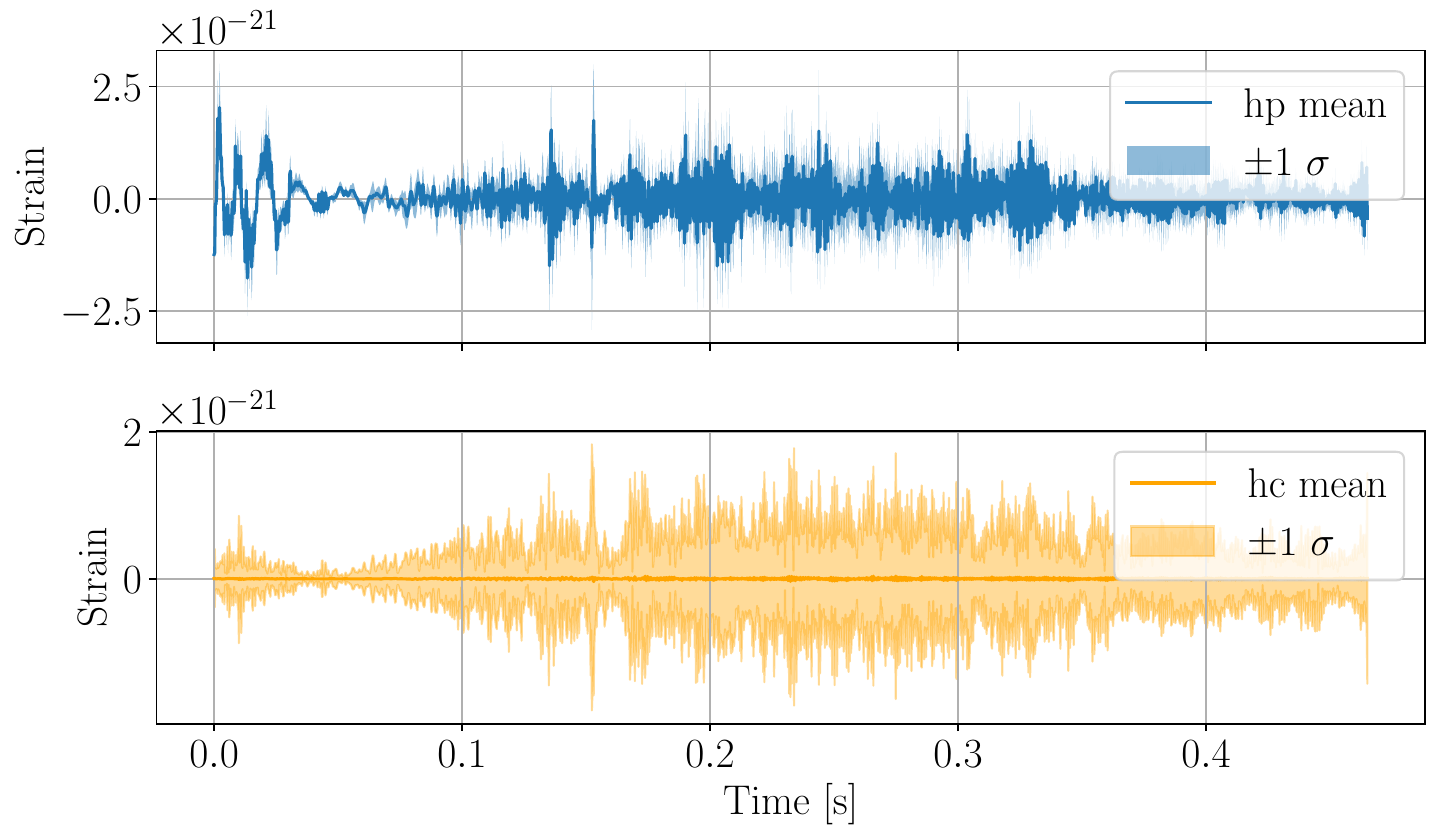}
    \includegraphics[width=\columnwidth]{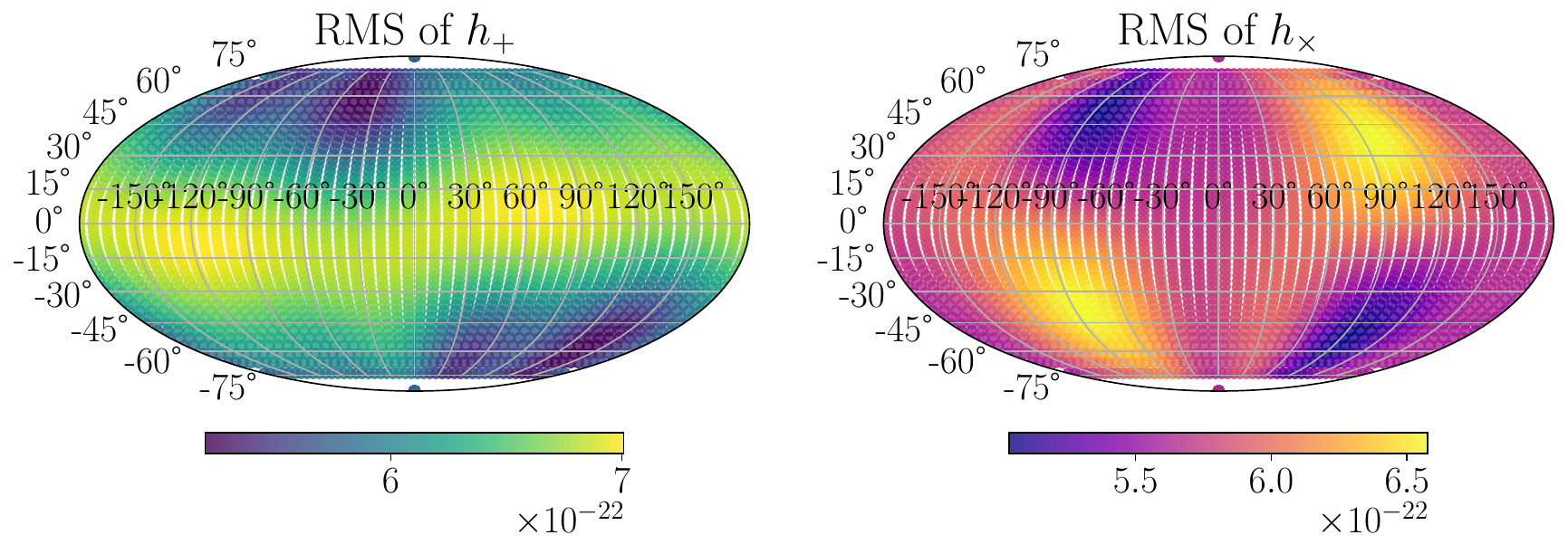}
    \caption{Mean and variance of gravitational-wave signals for the \texttt{s15\_SFHo\_nr} model~\cite{powell2024gravitational} at a source distance of 1~kpc, obtained by sampling $\cos\theta$ and $\phi$ uniformly.
    The bottom panel shows the angular dependence of the waveform amplitude.
    }
    \label{fig:hphc}
\end{figure}

Figure~\ref{fig:hphc} presents representative gravitational-wave signals from the CCSN simulations. The upper panel shows the mean and variance of the strain over the sampled source orientations, while the lower panel illustrates the angular dependence of the waveform amplitude. The mean value of $h_\times$ vanishes under spherical averaging over all source orientations, according to Eq.~(5)~\cite{szczepanczyk2021detecting}.

%m39\_b10, m39\_b12, rotation and neutrino and magnetic fields, \cite{powell2023three}
%s15\_sfho\_nr, s15\_sfhz\_nr\cite{powell2024gravitational}They all have magnetic fields and two are rotating, memory

\begin{table*}[t!]
\centering
\caption{Waveforms from 3D CCSN simulations used in this work.
Most waveform identifiers are taken from the catalog of Szczepańczyk et al.~\cite{szczepanczyk2021detecting}. References shown in the first column indicate additional 3D simulation waveforms incorporated in this study.
Models marked with $\dagger$ undergo a shock revival or explosion.
}
\label{tab:waveforms}
\begin{tabular}{llcccc}
\hline
Reference & Waveform Identifier & $M_{\rm ZAMS}$ $[M_\odot]$ &
$\Omega_c$ $[$rad/s$]$ & Magnetic Field & SASI\\
\hline

Andresen et al. (2017) & s11   & 11.2 &  - & - &-\\
                       & s20   & 20   &  - & - &(spiral) SASI\\
                       & s20s$\dagger$ & 20   &  - & - & SASI\\
                       & s27   & 27   &  - & - &SASI\\

Andresen et al. (2019) & m15fr$\dagger$ & 15 & 0.5  & Tayler-Spruit dynamo & (spiral) SASI\\
                       & m15nr & 15   &  -   & Tayler-Spruit dynamo & SASI\\
                       & m15r  & 15   & 0.2  & Tayler-Spruit dynamo & SASI\\

Kuroda et al. (2016)   & SFHx  & 15   & - & - &SASI\\
                       & TM1   & 15   & - & - &SASI\\

Kuroda et al. (2017)   & s11.2(SFHx) & 11.2 & - & - & - \\
                       & s15.0(SFHx) & 15   & - & - &SASI\\

Mezzacappa et al. (2020) & c15-3D$\dagger$ & 15 & - & - & SASI\\

Mezzacappa et al. (2023)~\cite{mezzacappa2023core}
                       & d\_15$\dagger$     & 15 & - & - & -\\
                       & d25\_3d$\dagger$   & 25 & - & - &SASI\\
                       & d9p6$\dagger$     & 9.6 & - & - &SASI\\

Müller et al. (2012)   & L15-3$\dagger$ & 15 & - & - &SASI\\
                       & N20-2$\dagger$ & 20 & - & - &SASI\\
                       & W15-4$\dagger$ & 15 & - & - &SASI\\

O'Connor \& Couch (2018) & mesa20      & 20 & - & - & SASI\\
                         & mesa20\_LR  & 20 & - & - & SASI\\
                         & mesa20\_pert& 20 & - & - & SASI\\
                         & mesa20\_pert\_LR & 20 & - & - & - \\
                         & mesa20\_v\_LR & 20 & - & - & -\\

Ott et al. (2013)      & s27-fheat1.00 & 27 & - & - &SASI\\
                       & s27-fheat1.05 & 27 & - & - &SASI\\
                       & s27-fheat1.10$\dagger$  & 27 & - & - &SASI\\
                       & s27-fheat1.15$\dagger$  & 27 & - & - &SASI\\

Powell \& Müller (2019) & s3.5\_pns$\dagger$ & 3.5 (non-ZAMS) & - & - & -\\
                        & s18$\dagger$       & 18  & - & - &-\\

Powell \& Müller (2020) & s18np & 18 & 3.4 & - & (spiral) SASI\\
                        & m39$\dagger$   & 39 & 0.54 & - &-\\
                        & y20$\dagger$   & 20 & - & - & -\\

Powell et al. (2021)~\cite{powell2021final}
                       & z85$\dagger$  & 85  & - & - & SASI\\
                       & z100\_sfho    & 100 &  - & - & SASI\\

Powell et al. (2023)~\cite{powell2023three}
                       & m39\_b10 & 39 &  0.54  & Dipole field ($10^{10}$G) & -\\
                       & m39\_b12 & 39 &  0.54  & Dipole field ($10^{12}$G) & - \\

Powell et al. (2024)~\cite{powell2024gravitational}
                       & s15\_sfho\_nr$\dagger$ & 15 &   -   & Dipole field ($10^{8}$G) & SASI\\
                       & s15\_sfhx\_nr$\dagger$ & 15 &   -   & Dipole field ($10^{8}$G) & SASI\\
Radice et al. (2019)   & s9$\dagger$  & 9  &  -  & - & -\\
                       & s10$\dagger$ & 10 &  -  & - & -\\
                       & s11$\dagger$ & 11 &  -  & - & -\\
                       & s12$\dagger$ & 12 &  -  & - & -\\
                       & s13 & 13 &  -  & - & SASI \\
                       & s14 & 14 & - & - & -\\
                       & s15 & 15 & - & - & -\\
                       & s19$\dagger$ & 19 & - & - & -\\
                       & s25$\dagger$ & 25 &  - & - & SASI\\
                       & s60$\dagger$ & 60 & - & - & -\\

Scheidegger et al. (2010) & R1E1CA\_L & 15 & 0.3 & \shortstack[l]{Toroidal field ($5\times10^{9}$G)\\Poloidal field($10^{6}$G)}& -\\
                           & R3E1AC\_L & 15 & 6.3 & \shortstack[l]{Toroidal field ($10^{6}$G)\\Poloidal field($5\times10^{9}$G)} & -\\
                           & R4E1FC\_L & 15 & 9.4 & \shortstack[l]{Toroidal field ($10^{12}$G)\\Poloidal field($5\times10^{9}$G)} & -\\

\hline
\end{tabular}
\end{table*}

\subsection{Detector configuration}
The Einstein Telescope (ET) is a proposed third-generation ground-based gravitational-wave detector.
We simulate its three interferometers, denoted as ET1, ET2, and ET3, arranged in a triangular configuration with an arm length of 10 km each.
For a source at right ascension $\alpha$, declination $\delta$, and polarization angle $\psi$, the projected signal strain in a detector $h(t)$ is
\begin{equation}
\begin{split}
h(t)
&= F_{+}(\alpha,\delta,\psi, t_{\mathrm{gps}})\,
   h_{+}(t-\tau; \theta,\phi)\\
\quad &+ F_{\times}(\alpha,\delta,\psi, t_{\mathrm{gps}})\,
   h_{\times}(t-\tau; \theta,\phi),
\end{split}
\end{equation}
where $F_{+,\times}$ are antenna-pattern functions, $t_{\mathrm{gps}}$ is GPS time, and $\tau$ is the geocentric time delay between the detector and the center of the Earth~\cite{Jaranowski_1998}.
Although the PyCBC library~\cite{pycbc} accounts for the $\tau$ delay, the maximum propagation delay between the ET interferometers ($\sim 33\,\mu\mathrm{s}$) is much smaller than the sampling interval ($\sim 244\,\mu\mathrm{s}$ at 4096~Hz), resulting in a negligible impact on the projected detector strain.
The GPS time is set to $t_{\mathrm{gps}} = 1577491218$ (01 Jan 2030, UTC).
The source orientation angles ($\theta, \phi$), and the right ascension, declination, and polarization angle ($\alpha,\delta, \psi$) are randomly assigned for each simulated waveform.
The signal strength is characterized by the network signal-to-noise ratio (SNR) $\rho_{\mathrm{network}}$, following the standard definition in Ref.~\cite{Abbott2016}:
\begin{equation}
\begin{aligned}
(a|b) &= 4 \mathrm{Re}\int_{f_{\mathrm{min}}}^{f_{\mathrm{max}}}
\frac{\tilde{a}(f)\tilde{b}^*(f)}{S(f)}\,\mathrm{d}f,\\
\rho_{\mathrm{network}} &= \sqrt{\sum_{j=1}^{D_{\rm{det}}}(h_j|h_j)}.
\end{aligned}
\label{eq:inner_product_snr}
\end{equation}
Here, $D_{\rm{det}}=3$ is the number of detectors, $h_j$ denotes the injected waveform in $j$th detector, and $(a|b)$ represents the noise-weighted inner product defined using the one-sided noise power spectral density $S(f)$. The integration limits are set to $f_{\mathrm{min}}=100$ Hz and $f_{\mathrm{max}}=1200$ Hz, covering the dominant frequency range of gravitational waves expected from CCSNe. For reference, signals with a network SNR of approximately $10$ or higher are considered to have the potential to be detected by burst-search pipelines~\cite{szczepanczyk2021detecting}.

\subsection{Signal preprocessing}
CCSN gravitational-wave signals are expected to contain a low-frequency memory effect~\cite{muller2004toward, kotake2009stochastic, richardson2024detecting, powell2024gravitational}, which is concentrated below a few tens of hertz and can enhance the detectability of CCSNe. 
In particular, the memory lies predominantly outside
the sensitive frequency band of current ground-based detectors and,
although it can be calculated from numerical simulations, is sometimes not included in the GW waveforms.
In this study, we apply a band-pass filter between 100 and 1200~Hz to focus on the oscillatory components in the analysis band.
%Since the DC component is not represented in conventional time--frequency analysis methods such as the short-time Fourier transform (STFT) and its extension, the wavelet transform, and because detector sensitivity is limited at such low frequencies, we apply a Butterworth band-pass filter between 100 and 1200~Hz. This allows us to focus on the dominant oscillatory components of CCSN gravitational-wave signals.

\begin{figure}[h!]
    \centering
    \includegraphics[width=\columnwidth]{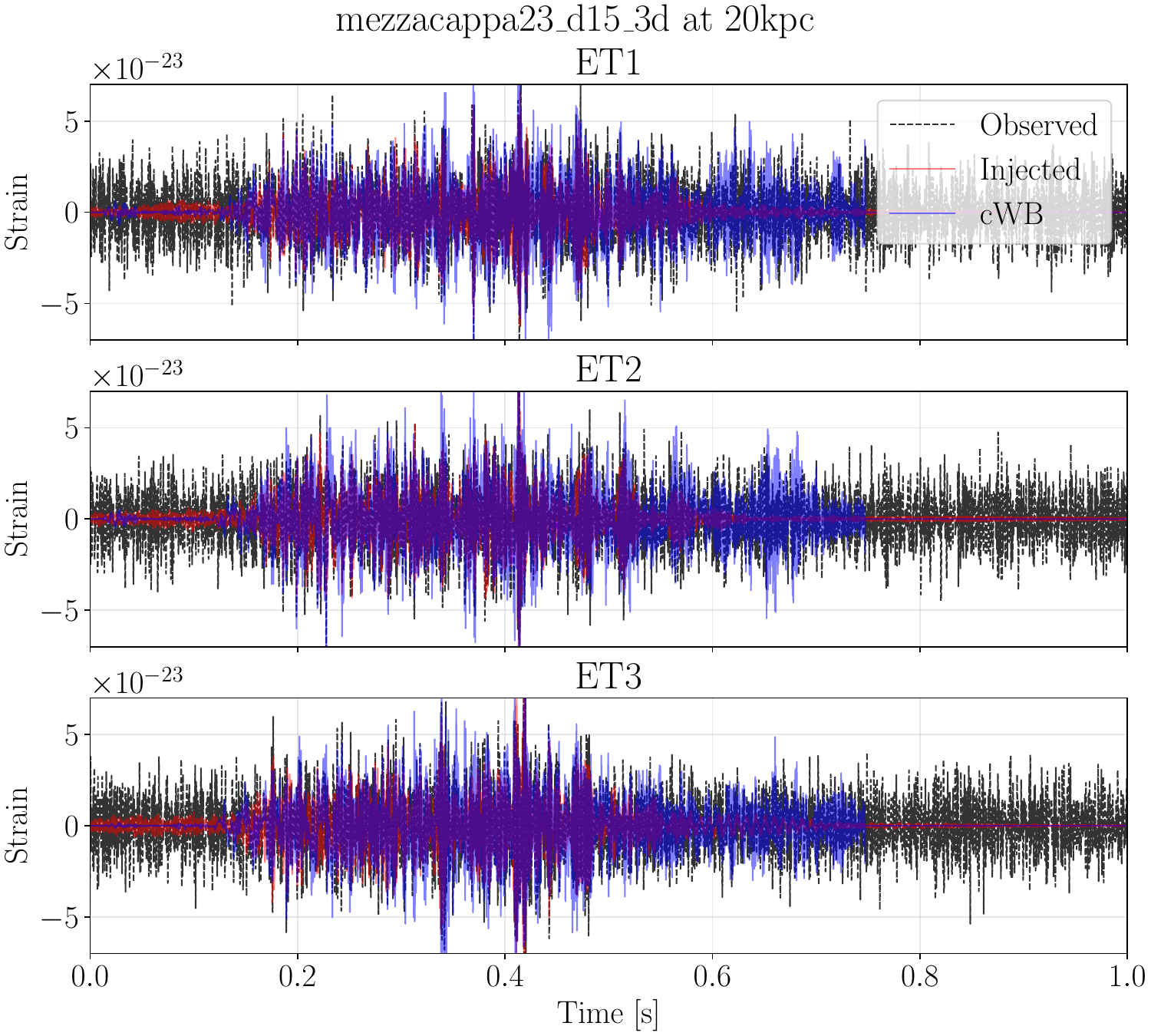}
    \caption{Representative cWB reconstruction for the \texttt{d15\_3d} model~\cite{mezzacappa2023core} at a source distance of 20~kpc. The observed strain with noise (black dashed), the injected waveform (red), and the cWB-reconstructed waveform (blue) are shown for the three detectors (ET1--ET3). The source orientation $(\theta,\phi)$ and the detector antenna pattern $(\alpha,\delta,\psi)$ are randomly selected.
    }
    \label{fig:mezzacappa23_d15_3d}
\end{figure}

Independent colored Gaussian noise generated from the assumed ET power spectral density is added to the projected signal in each interferometer. The detector data are processed using coherent WaveBurst (cWB)~\cite{Klimenko2008,Klimenko2016,Drago2021}. cWB is a configurable, model-independent search pipeline designed to detect and reconstruct transient GW signals from data collected by a global network of GW detectors.
Its algorithm is based on a time--frequency analysis using the Wilson--Daubechies--Meyer wavelet transform, enabling the identification and coherent reconstruction of GW transients without prior knowledge of the signal waveform.
Figure~\ref{fig:mezzacappa23_d15_3d} shows a representative result of cWB reconstruction. The reconstructed waveform captures the main features of the injected CCSN gravitational-wave signal despite the presence of detector noise.

The cWB pipeline is optional. When it is not applied, each detector time series is whitened using the theoretical ET sensitivity curve.

\section{Time--frequency representations}\label{sec3}
Time--frequency representations often provide more informative representations for deep-learning models than raw time-series data~\cite{10.1145/3711896.3736571}.
Motivated by these findings, the present study employs time--frequency representations as the input to the deep-learning models.

\subsection{Short-Time Fourier Transform}
\begin{figure*}[htb!]
    \centering
    \includegraphics[width=2\columnwidth]{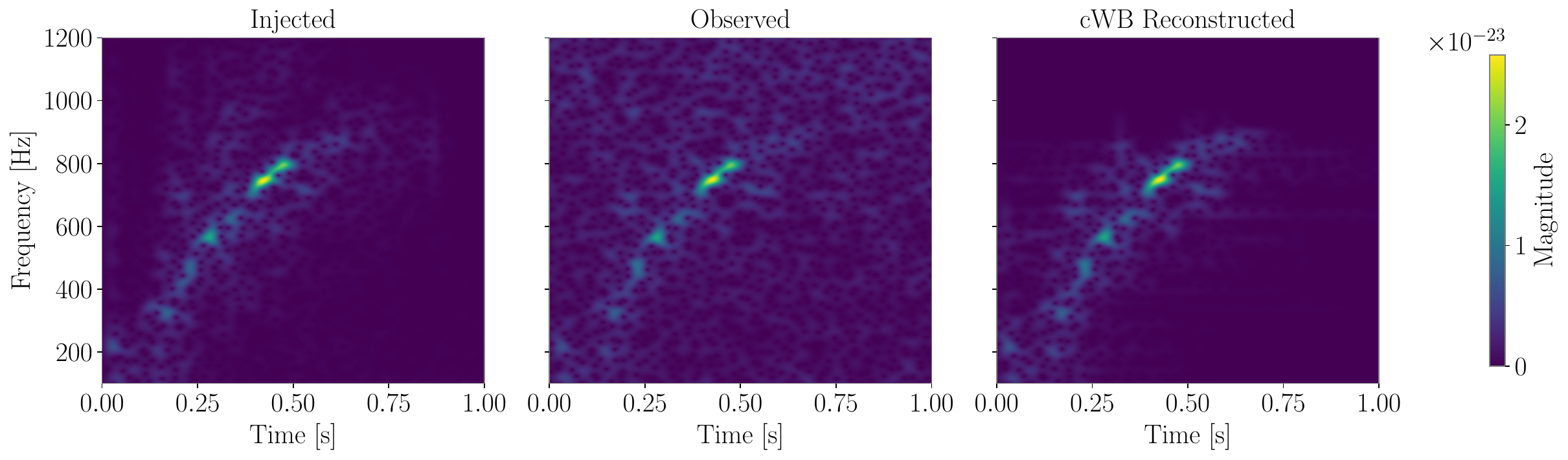}
    \caption{STFT spectrogram for the \texttt{m39} model at a source distance of 50~kpc. The spectrogram is obtained from the ET1 data. The network SNR computed from the three detectors is 144.
    }
    \label{fig:m39_STFT}
\end{figure*}

The waveforms are subsequently transformed into time--frequency representations for waveform classification.
As a standard time--frequency analysis method, we employ the short-time Fourier transform (STFT).
Figure~\ref{fig:m39_STFT} shows the STFT spectrograms of the injected (left), observed (middle), and cWB-reconstructed (right) signals. The cWB reconstruction effectively suppresses the background noise while preserving the characteristic spectral features of the CCSNe.

\subsection{Hilbert--Huang Transform}
Alternatively, we apply the Hilbert--Huang transform (HHT)~\cite{Huang1998}, an advanced adaptive time--frequency analysis technique, and compare the resulting classifier with the STFT-based classifier.
The HHT decomposes a signal into intrinsic mode functions (IMFs) using empirical mode decomposition (EMD), followed by Hilbert spectral analysis to obtain a high-resolution time--frequency representation. 
The EMD decomposes an input signal $x(t)$ as
\begin{equation}
x(t)=\sum_{j=1}^{J}c_j(t)+r(t),
\end{equation}
where $c_j(t)$ denotes the $j$th intrinsic mode function (IMF), $J$ is the number of IMFs, and $r(t)$ is the residual component. The corresponding analytic signal is defined as
\begin{equation}
z_j(t)=c_j(t)+i\mathcal{H}[c_j](t)=a_j(t)e^{i\phi_j(t)},
\end{equation}
where $\mathcal{H}[\cdot]$ denotes the Hilbert transform. The instantaneous amplitude (IA) and instantaneous frequency (IF) are given by $a_j(t)=|z_j(t)|$ and $f_j(t)=(2\pi)^{-1}d\phi_j/dt$, respectively.
Basic EMD is susceptible to decomposition artifacts such as \textit{mode shifting} and \textit{mode splitting}~\cite{Wu2009}.
Mode shifting occurs when a time--frequency feature is represented by a different IMF, whereas mode splitting distributes a single time--frequency feature across multiple IMFs.
Ensemble EMD~\cite{Wu2009} suppresses these artifacts by adding white noise to the input signal and averaging the resulting ensemble of decompositions. Complementary ensemble EMD (CEEMD)~\cite{Yeh2010} guarantees the exact reconstruction of the original signal from the superposition of the decomposed IMFs despite the addition of white noise.

\begin{figure}[h]
    \centering
    \includegraphics[width=\columnwidth]{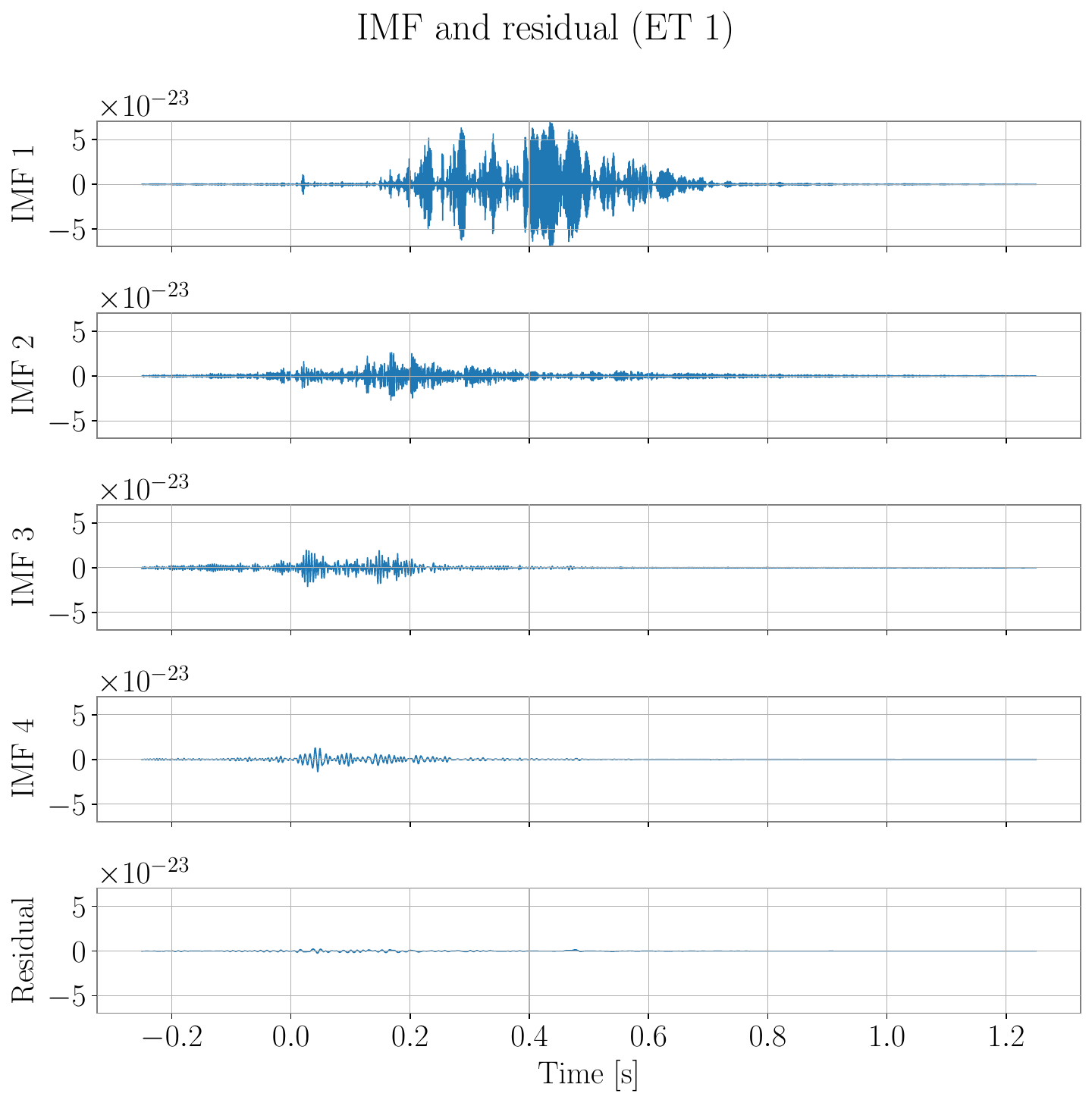}
    \includegraphics[width=\columnwidth]{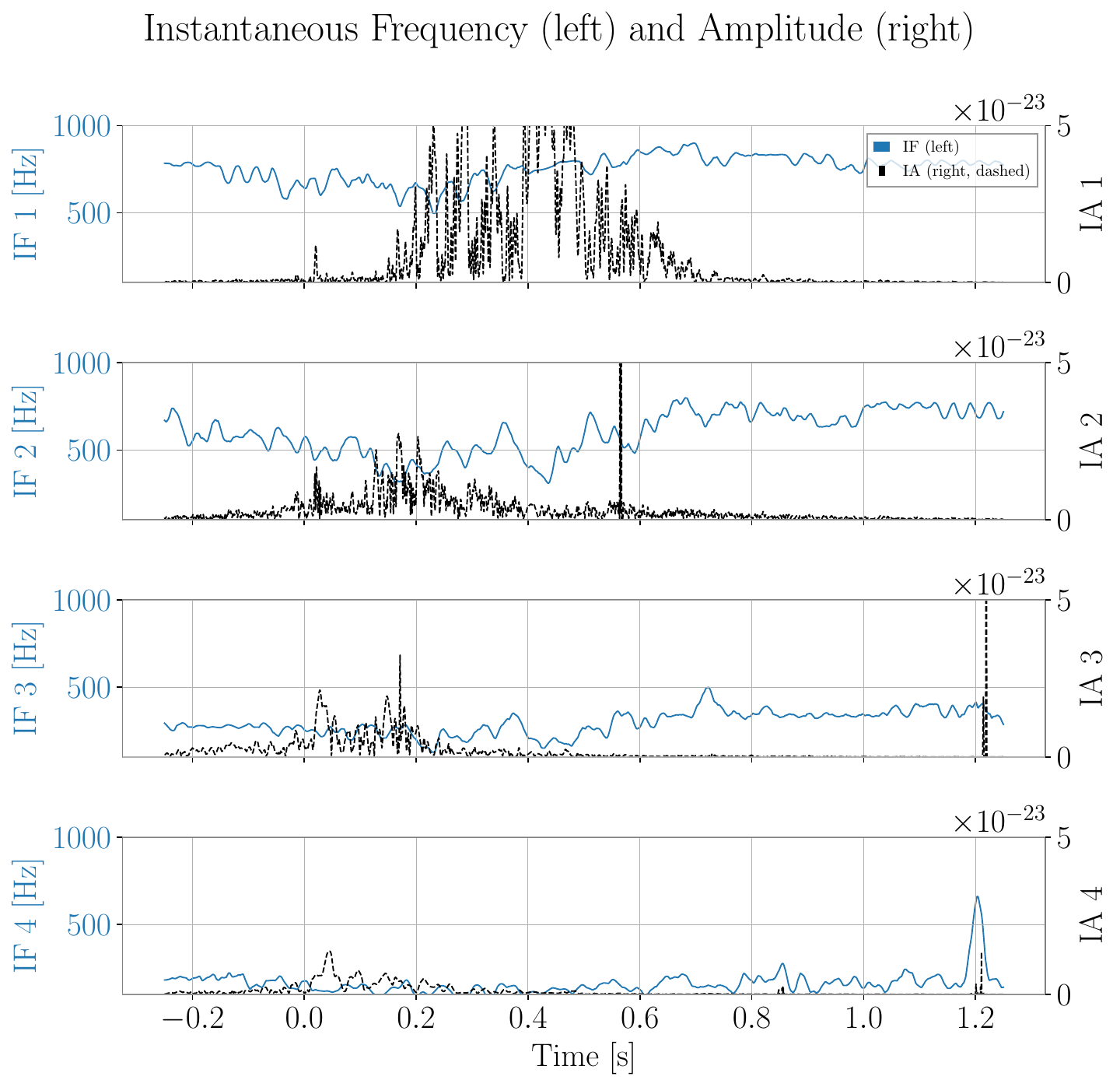}
    \caption{The IMFs and residual (top), and the corresponding IF and IA (bottom), obtained by HHT for the \texttt{m39} model at a source distance of 50~kpc. The input waveform is the cWB-reconstructed signal shown in Fig.~\ref{fig:m39_STFT}.
    A visualized spectrogram is also given in Appendix~\ref{appendix:hht}.
    }
    \label{fig:m39_HHT}
\end{figure}

The CEEMD configuration we use follows Ref.~\cite{PhysRevD.110.104020} and is applied uniformly to all waveforms, with the number of IMFs fixed at four. The HHT decomposes the input signal into IMFs in descending order of frequency. Consequently, the last IMF and its corresponding IA and IF contain the lowest-frequency components. Figure~\ref{fig:m39_HHT} shows representative IMFs together with their corresponding IFs and IAs obtained from the \texttt{m39} model. Typical CCSN GW signals exhibit frequency features extending up to approximately 1~kHz, and their signal power can therefore be distributed across multiple IMFs.

A given CCSN feature is not necessarily assigned to a specific IMF. Its IMF assignment may vary depending on the waveform morphology, detector noise, and the EMD/CEEMD configuration.
Moreover, the HHT may exhibit artifacts such as \emph{mode shifting} and \emph{mode splitting}.
Although these phenomena complicate the interpretation of individual IMFs, the underlying time--frequency information is generally preserved across the decomposed components.
This suggests that the application of deep learning may enable robust feature learning by integrating information distributed across multiple IMFs, IFs, and IAs rather than relying on individual components.
Therefore, the time-series representations of the IMFs together with their corresponding IFs and IAs, which preserve the essential time--frequency information, are used as the input to the axial attention module. The resulting feature representations are then embedded as tokens for subsequent transformer layers.
The architecture is described in Sec.~\ref{subsubsec:aat}.

\section{Deep Learning}\label{sec4}

\subsection{Dataset}
\begin{figure*}[htb!]
    \centering
    \includegraphics[width=1.9\columnwidth]{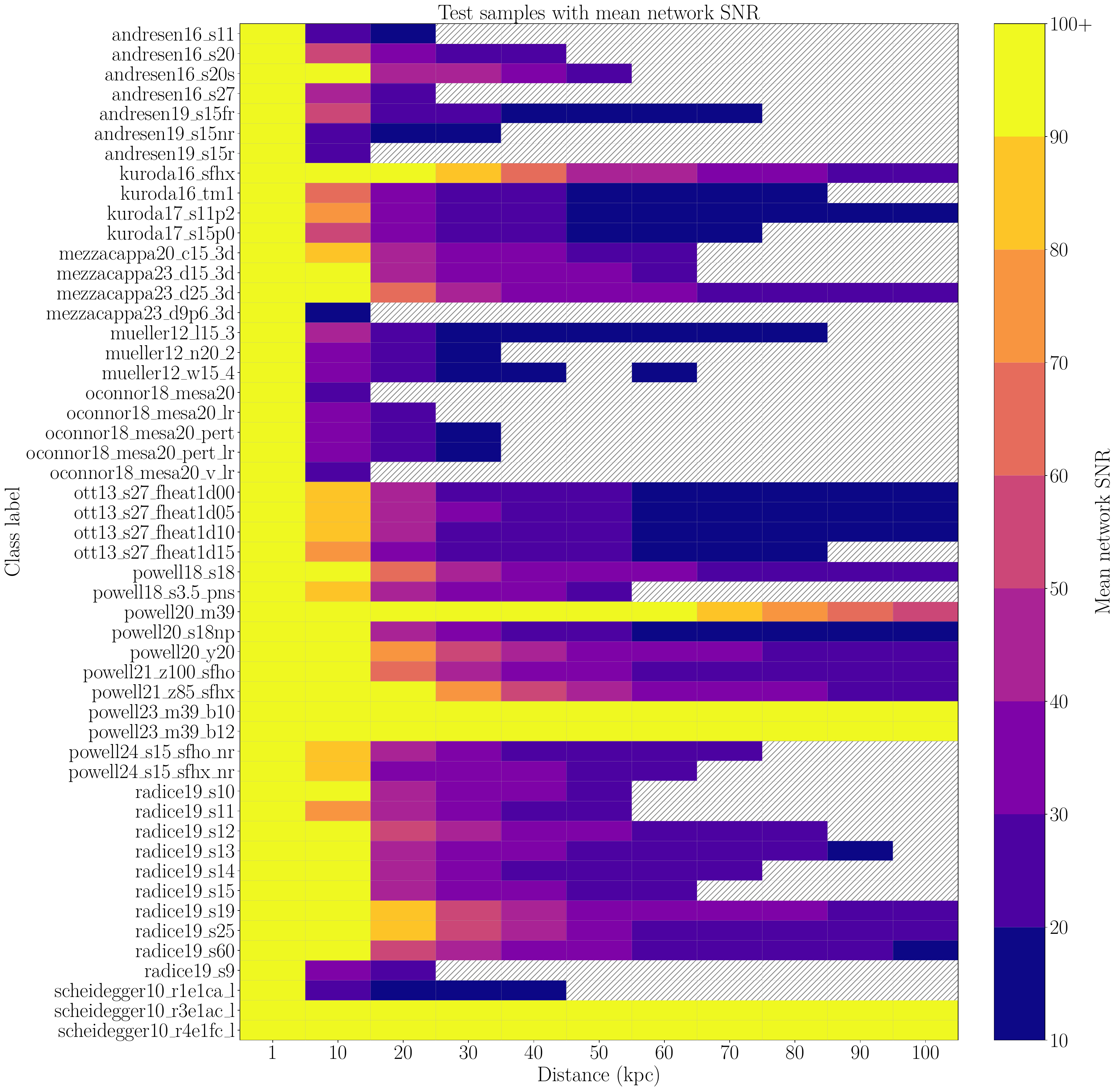}
    \caption{Distribution of the 51 CCSN waveforms after processing with the available cWB pipeline at each source distance. Colors indicate the mean network SNR of the available events in each class--distance bin. Hatched cells indicate that no reconstructed events are available, while the remaining events constitute the test dataset used for the classification task. The source orientation $(\theta,\phi)$ and the detector antenna pattern $(\alpha,\delta,\psi)$ are randomly selected.
    }
    \label{fig:dataset}
\end{figure*}

We sampled 50 azimuthal angles uniformly in $\phi\in[0,2\pi]$ and 50 polar angles uniformly in $\cos\theta\in[-1,1]$, resulting in 2,500 source orientations for each CCSN waveform. Assuming observations with the ET, the detector antenna pattern was randomly sampled for each realization. For each waveform, simulated detector strains were generated at source distances of 1~kpc  and 10--100~kpc in steps of 10~kpc. The maximum distance of 100~kpc was motivated by previous estimates of the CCSN detection for ET~\cite{ccsn-et}.
The same number of events was generated at each distance to evaluate how reconstruction availability and classification performance vary with source distance, without accounting for the expected astrophysical event rate at each distance.
With 51 waveform models, this procedure yielded a total of $2,500 \times 11 \times 51 = 1,402,500$ samples before accounting for missing records and cWB reconstruction availability. Figure~\ref{fig:dataset} shows the resulting training and validation sample counts after selecting available cWB reconstructions. The counts depend strongly on waveform class and distance, so this selection introduces class imbalance. The available training and validation realizations are split in an 80:20 ratio, stratified by waveform class. For evaluation, an independent test dataset, separate from the training and validation data, was generated using 100 new samples for each waveform model at each source distance.

\subsection{Network Architectures}
For the STFT representation, we employ ConvNeXt V2 because the STFT produces a two-dimensional time--frequency image, allowing the model to effectively extract local spatial patterns through convolutional operations. In contrast, the HHT representation consists of multiple synchronized time-series (IMFs, IFs, and IAs) rather than a single image. Therefore, we adopt an axial attention transformer for HHT input, which is well suited for modeling long-range temporal dependencies and correlations across multiple channels.

\subsubsection{ConvNext V2}
In image recognition, ResNet~\cite{resnet} was a breakthrough convolutional neural network (CNN) architecture that introduced residual connections and batch normalization, significantly improving image classification performance. Subsequently, the Vision Transformer~\cite{vit}, a Transformer-based image recognition model, achieved even better performance. ConvNeXt~\cite{convnext, convnextv2} revisits the CNN architecture by incorporating design principles inspired by Transformers while retaining convolutional operations, achieving state-of-the-art performance comparable to Transformer-based models. Previous studies on CCSN gravitational-wave classification have predominantly employed CNN-based architectures~\cite{Iess_2020, PhysRevD.102.043022, PhysRevD.108.123033}. Therefore, as a baseline, we adopt the ConvNeXt~V2 architecture, using STFT spectrograms as the input representation.

ConvNeXt modernizes CNNs by incorporating several design principles inspired by Vision Transformers, such as large-kernel depthwise convolutions, layer normalization, and GELU activation. ConvNeXt~V2 further improves the architecture by introducing Global Response Normalization (GRN).
Although a fully convolutional masked autoencoder (FCMAE) was also proposed in the original paper as a self-supervised pretraining strategy, one of the objectives of this study is to evaluate the effectiveness of the STFT and HHT representations. Therefore, our model is trained using a standard supervised learning procedure without FCMAE pretraining.
Appendix~\ref{appendix:cnn} provides details of the architecture used here.

\subsubsection{Axial Attention Transformer}\label{subsubsec:aat}
Attention~\cite{transformer} is a mechanism that models dependencies between input features by computing weighted interactions through query ($Q$), key ($K$), and value ($V$) representations.
Let $X\in\mathbb{R}^{N\times D}$ denote the input feature matrix, where $N$ is the number of input tokens and $D$ is the embedding dimension.
The query, key, and value matrices,
$Q\in\mathbb{R}^{N\times d}$,
$K\in\mathbb{R}^{N\times d}$, and
$V\in\mathbb{R}^{N\times d}$,
are obtained by
\begin{equation}
Q = XW_Q, \quad K= XW_K, \quad V = XW_V,    
\end{equation}
where $W_Q$, $W_K$, and $W_V$ are learnable projection matrices in $\mathbb{R}^{D\times d}$, and $d$ is the embedding dimension of the query, key, and value representations.
The attention weight matrix is defined as
\begin{equation}
A
=
\mathrm{softmax}
\left(
\frac{QK^{\mathsf T}}{\sqrt{d}}
\right),
\label{eq:attention_weights}
\end{equation}
where $d$ is the dimension of the query and key vectors.
The attention output is then given by
\begin{equation}
\mathrm{Attention}(Q,K,V)
=
AV.
\label{eq:attention}
\end{equation}
However, its computational cost scales quadratically with the number of input tokens $N$, i.e., $\mathcal{O}(N^2)=\mathcal{O}((SC)^2)$ where $N=S\times C$, $S\in\mathbb{N}$ is the sequence length, and $C\in\mathbb{N}$ is the number of channels.
Axial attention~\cite{ho2019axial} addresses this limitation by applying self-attention independently along each axis of a multidimensional tensor, reducing the computational cost to $\mathcal{O}(CS^2+SC^2)$. For $S\gg C$, this scales as $\sim \mathcal{O}(CS^2)$ while maintaining a global receptive field through stacked axial attention layers.

Since HHT representations typically have a large sequence length $S$, corresponding to a high temporal resolution, axial attention is employed to reduce computational costs.
The IMF, instantaneous frequency (IF), and instantaneous amplitude (IA) representations are first processed by separate token streams with shared axial-attention blocks.
Within each encoder, axial self-attention is sequentially applied along the mode, temporal-patch, and detector axes.
Mode-axis attention captures relationships among intrinsic mode functions within the same HHT representation, temporal-axis attention models long-range dependencies across temporal patches, and detector-axis attention captures coherent responses across detectors.
Since the IMF, IF, and IA token streams remain separate throughout the hierarchical axial stages, a latent mixer layer is introduced to integrate information from all token embeddings through cross attention.
In the cross-attention operation defined by Eq.~\eqref{eq:attention}, the learnable latent tokens serve as the queries ($Q=Q_{\mathrm{latent}}$), whereas the structured IMF, IF, and IA token embeddings serve as the keys ($K$) and values ($V$).
The resulting aggregated representations are subsequently used for classification.
Appendix~\ref{appendix:axial} provides details of the architecture used here.

%HHT data use a custom axial transformer.  A temporal unfolding operation creates overlapping patches of length 64 with stride 32.  Each component and time-patch token is projected to 128 dimensions.  When detector attention is enabled, self-attention first mixes the three detector features within each component--patch cell.  Four subsequent axial-attention blocks, each with four heads and a 256-dimensional feed-forward layer, exchange information along the component and time axes.  Learned component and time embeddings retain the identity and order of tokens; mean pooling followed by a linear layer produces class logits.  This factorization is computationally more economical than dense attention over all component--time--detector samples and encodes the natural axes of the HHT representation.

\subsection{Training strategies}
We consider two complementary analysis approaches for both the STFT- and HHT-based models. The first follows a conventional analysis pipeline, in which the models are trained on signals reconstructed by cWB. The second adopts a direct deep-learning approach, in which the initially trained models are fine-tuned on whitened detector data without any reconstructions, enabling the classification of signals directly from the detector data.

\subsubsection{Initial training with cWB-reconstructed signals}

Curriculum learning~\cite{10.1145/1553374.1553380} is employed: training begins with events at a source distance of 1~kpc and progressively introduces samples at distances from 10 to 100~kpc over 110 epochs.
The STFT and HHT models are trained for 110 epochs.
The models are optimized using the SAM~\cite{DBLP:conf/iclr/ForetKMN21} optimizer with AdamW as the base optimizer. Cross-entropy loss is used as the objective function. The initial learning rates are $10^{-3}$ for STFT and $10^{-4}$ for HHT, and the SAM neighborhood size is $0.03$. A \textit{ReduceLROnPlateau} learning-rate scheduler is employed with a reduction factor of 0.7 and a patience of 5 epochs. 
The input signal spans the time interval from $-0.25$ to 1.25~s relative to the onset of the CCSN gravitational-wave signal, where $t=0$ corresponds to the signal onset. During training, time-shift augmentation is applied by randomly shifting the cropping window by up to $\pm$0.1~s within this interval. This augmentation improves robustness against variations in the signal arrival time.
Training was performed using four RTX 4500 Ada Generation GPUs with 24~GB of memory each, hosted on a server equipped with an Intel\textsuperscript{\textregistered} Xeon\textsuperscript{\textregistered} Gold 6526Y CPU (32 cores, 2.8~GHz).

\subsubsection{Fine-tuning with unreconstructed signals and noise}
For raw-input classification, each model is initialized from its cWB-trained counterpart and fine-tuned with a newly initialized 52-class (51 signal classes and one noise class) output layer.
Although the raw detector data differ from cWB-reconstructed signals because they retain detector noise and have not undergone cWB reconstruction, the underlying CCSN signals share the same characteristic time--frequency structures.
We therefore use the cWB-trained models as initialization to retain the learned representations of CCSN waveform features, while adapting them to the raw-input distribution and avoiding the need to train the models from scratch.

The fine-tuning dataset contains whitened detector data from both reconstruction-available and reconstruction-unavailable events, selecting up to 200 events from each category per class and distance, together with noise-only data. Both models are fine-tuned for 110 epochs, with initial learning rates of $10^{-4}$ for STFT and $8\times10^{-5}$ for HHT. The test dataset is generated with the same configuration, with 100 events per waveform and source distance.

\subsubsection{Evaluation metrics}

We evaluate classification performance using accuracy, macro-$F_1$, and class-wise $F_1$ scores.
Accuracy is the fraction of correctly classified events.
The class-wise $F_1$ score for class $i$ is defined as
\begin{equation}
F_{1,i}=\frac{2\,\mathrm{TP}_i}{2\,\mathrm{TP}_i+\mathrm{FP}_i+\mathrm{FN}_i},
\label{eq:f1}
\end{equation}
where $\mathrm{TP}_i$, $\mathrm{FP}_i$, and $\mathrm{FN}_i$ are the true-positive, false-positive, and false-negative counts for class $i$.
Class-wise $F_1$ scores reveal performance differences among individual classes, while macro-$F_1$ gives equal weight to each class regardless of the number of samples.
The macro-$F_1$ score is defined as
\begin{equation}
F_1^{\rm macro}=\frac{1}{N_{\mathrm{class}}}\sum_{i=1}^{N_{\mathrm{class}}}F_{1,i},
\end{equation}
where $N_{\mathrm{class}}$ is the number of classes.

To evaluate the consistency between the STFT- and HHT-based classifiers, we establish a one-to-one correspondence between common test events using their event identifiers, ground-truth labels, and source distances.
Each event is randomly generated and individually tracked, while the subsequent processing differs between the cWB-reconstructed and raw inputs and between the STFT and HHT representations.
Joint accuracy is defined as the fraction of common test events for which both the STFT- and HHT-based classifiers predict the correct label:
\begin{equation}
\mathrm{Joint\ accuracy}
=
\frac{N(\mathrm{STFT\ correct} \cap \mathrm{HHT\ correct})}
     {N(\mathrm{common\ events})}.
\label{eq:joint_accuracy}
\end{equation}
For example, if two classifiers have an accuracy of $0.90$, their expected joint accuracy is $0.81$ under the assumption of independent classification outcomes.

We define prediction confidence as the largest softmax output probability and evaluate calibration using the expected calibration error (ECE).
The ECE summarizes the discrepancy between the prediction confidence and the empirical accuracy over $M$ confidence bins. 
\begin{equation}
\mathrm{ECE} = \sum_{m=1}^{M} \frac{|B_m|}{N_{\mathrm{eval}}}
    \left|
    \mathrm{acc}(B_m) - \mathrm{conf}(B_m)
    \right|,
\label{eq:ece}
\end{equation}
where $N_{\mathrm{eval}}$ is the total number of evaluated predictions, $B_m$ denotes the set of samples in the $m$th confidence bin, $\mathrm{acc}(B_m)$ is the empirical accuracy, and $\mathrm{conf}(B_m)$ is the mean predicted confidence within the bin. A smaller ECE indicates better agreement between prediction confidence and empirical accuracy.

\section{Result}\label{sec5}
%We evaluate two classification settings. In the first, STFT and HHT features are computed from cWB-reconstructed signals, and the task is to identify one of the 51 numerical waveform classes. In the second, the classifiers receive features computed directly from the whitened detector data, including events without an available cWB reconstruction. This second setting has 52 classes because it includes a noise-only class. Hereafter, ``raw input'' denotes this latter preprocessing choice; the neural networks still receive STFT or HHT features including detector noise rather than untransformed strain.

\subsection{Classification of cWB-reconstructed signals}\label{subsec:results_cwb}
\begin{figure*}[htb!]
    \centering
    \includegraphics[width=2.0\columnwidth]{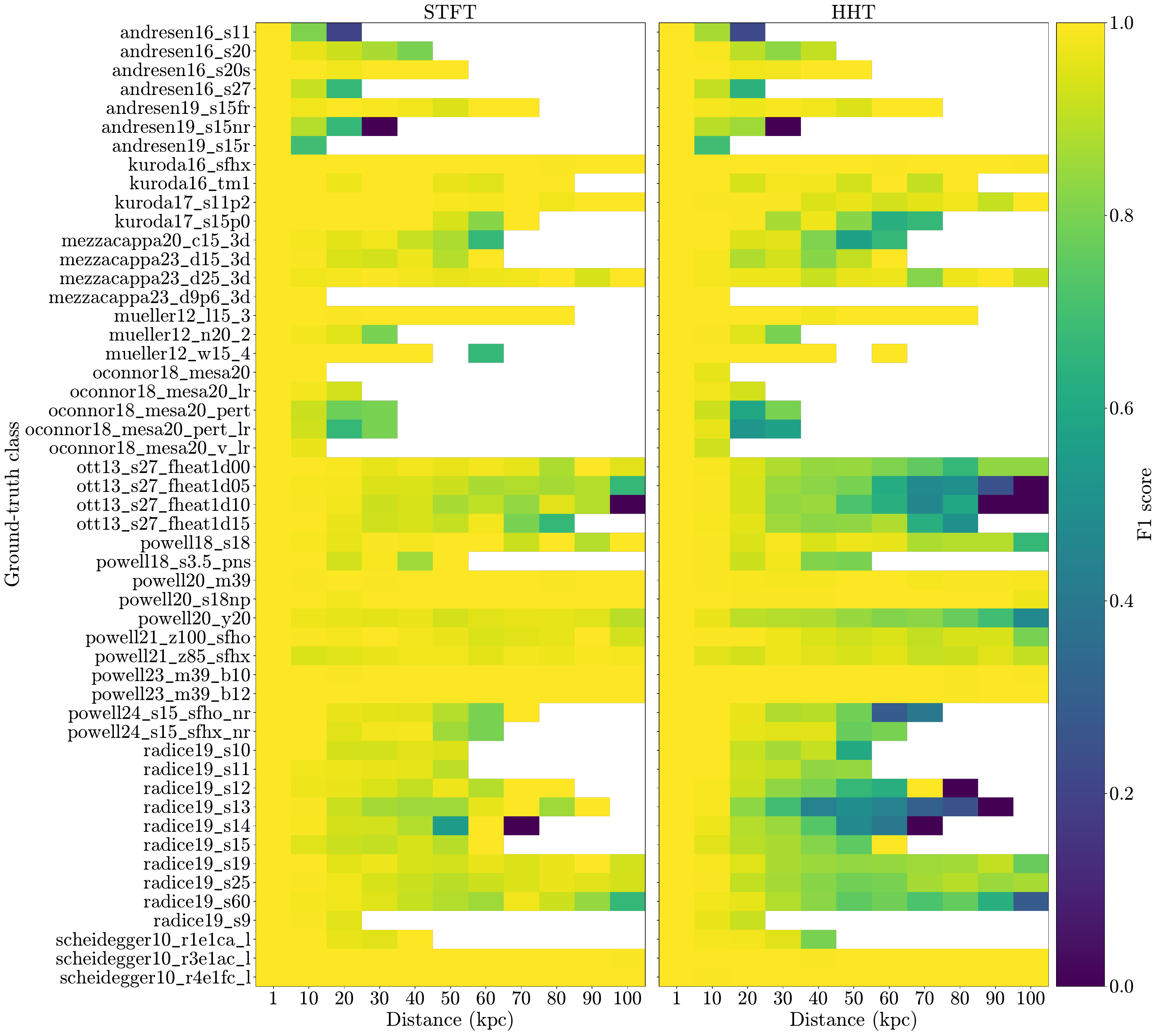}
    \caption{Class-wise $F_1$ scores for STFT (left) and HHT (right) on cWB-reconstructed test signals at each distance, without confidence selection. White cells indicate that no ground-truth test event of that class is available at that distance; they do not denote zero F1.
    }
    \label{fig:class_f1_distance}
\end{figure*}

We first assess whether the STFT- and HHT-based classifiers can distinguish the simulated waveform classes when a cWB reconstruction is available.
Table~\ref{tab:test_accuracy_by_distance} summarizes the dependence of accuracy on source distance. Both classifiers achieve 100\% accuracy at 1~kpc. At larger distances, however, the classification accuracy does not remain at 100\%, even when cWB reconstruction is successful. For STFT, the accuracy is approximately 96\% at 50~kpc and 99\% at 100~kpc. This apparent variation in performance is primarily due to changes in the set of waveform classes for which cWB reconstruction is available. All 51 classes are represented at 1 and 10~kpc, whereas the number decreases to 36 at 50~kpc and 19 at 100~kpc. Figure~\ref{fig:class_f1_distance} shows the corresponding class-wise $F_1$ scores as a function of source distance. At 50~kpc, some weak signals are still reconstructed but can be misclassified as other waveform classes. At 100~kpc, in contrast, the remaining classes are dominated by stronger signals. Thus, the apparent increase in accuracy to approximately 99\% at 100~kpc reflects the reduced and more readily reconstructable set of waveform classes rather than improved classification performance at larger distances.

%Both representations maintain high scores for several classes over a broad distance range, including \texttt{powell20\_m39}, \texttt{powell23\_m39\_b10}, and \texttt{powell23\_m39\_b12}. In contrast, the Ott et al.\ waveform family and several Radice et al.\ models show a stronger decline, particularly for HHT. 

\begin{table}
\caption{Classification accuracy on cWB-reconstructed test signals at each source distance. ``Classes'' is the number of distinct ground-truth waveform classes represented at that distance; the classifier retains 51 output classes. Values in parentheses use only predictions with confidence $\geq0.9$, selected independently for STFT and HHT.}
\label{tab:test_accuracy_by_distance}
\begin{tabular}{rrcc}
\toprule
Distance (kpc) & Classes & STFT & HHT \\
\midrule
1 & 51 & 100.0\% (100.0\%) & 100.0\% (100.0\%) \\
10 & 51 & 98.3\% (99.4\%) & 98.4\% (99.9\%) \\
20 & 47 & 96.8\% (99.0\%) & 94.7\% (99.4\%) \\
30 & 43 & 96.9\% (98.9\%) & 92.7\% (99.2\%) \\
40 & 39 & 96.8\% (99.3\%) & 90.8\% (99.7\%) \\
50 & 36 & 95.8\% (98.7\%) & 90.1\% (99.9\%) \\
60 & 33 & 97.2\% (99.2\%) & 91.5\% (99.1\%) \\
70 & 28 & 97.6\% (99.9\%) & 91.2\% (99.5\%) \\
80 & 24 & 98.1\% (99.9\%) & 94.0\% (99.7\%) \\
90 & 20 & 99.0\% (100.0\%) & 95.2\% (99.9\%) \\
100 & 19 & 98.8\% (99.9\%) & 95.2\% (100.0\%) \\
\bottomrule
\end{tabular}
\end{table}

\begin{figure*}[htb!]
    \centering
    \begin{subfigure}{\textwidth}
        \centering
        \includegraphics[width=\textwidth]{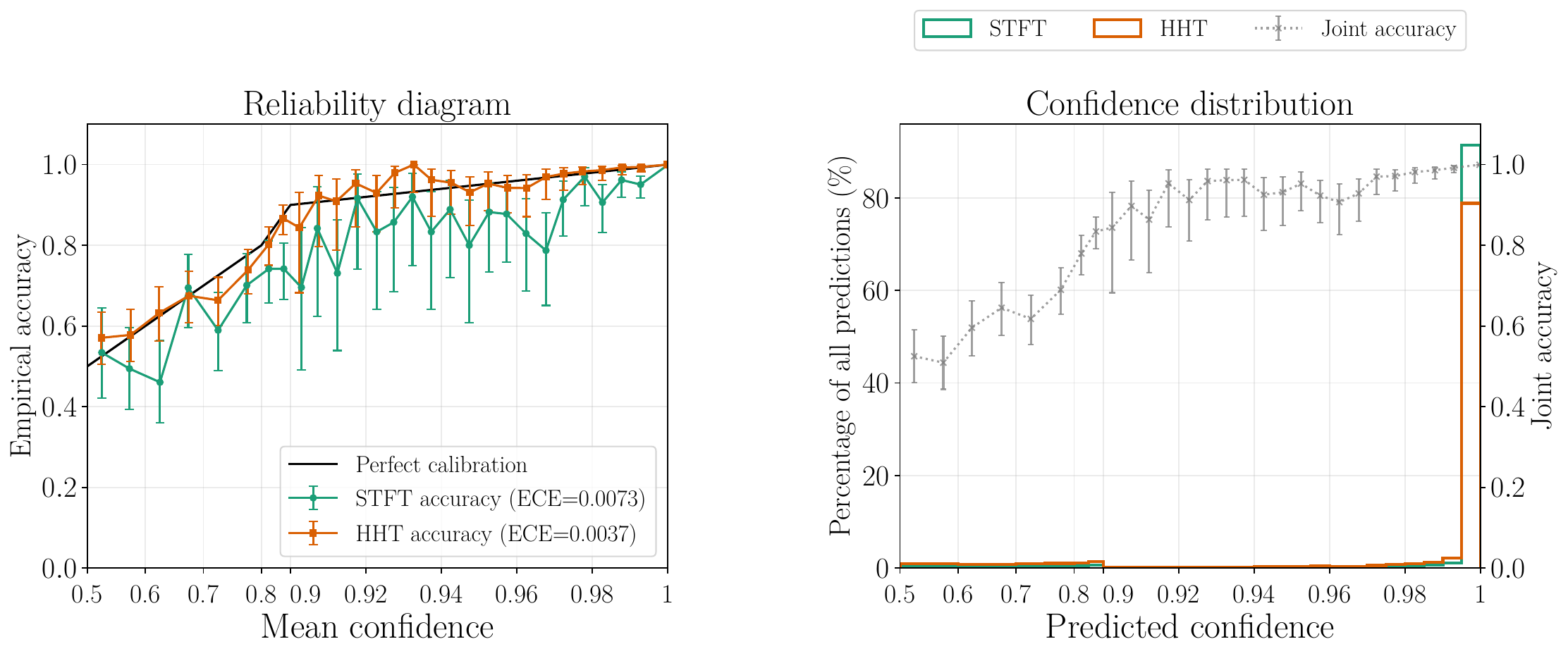}
        \caption{Reconstruction available (23,500 samples)}
        \label{fig:ece-cwb-available}
    \end{subfigure}

    \par\medskip

    \begin{subfigure}{\textwidth}
        \centering
        \includegraphics[width=\textwidth]{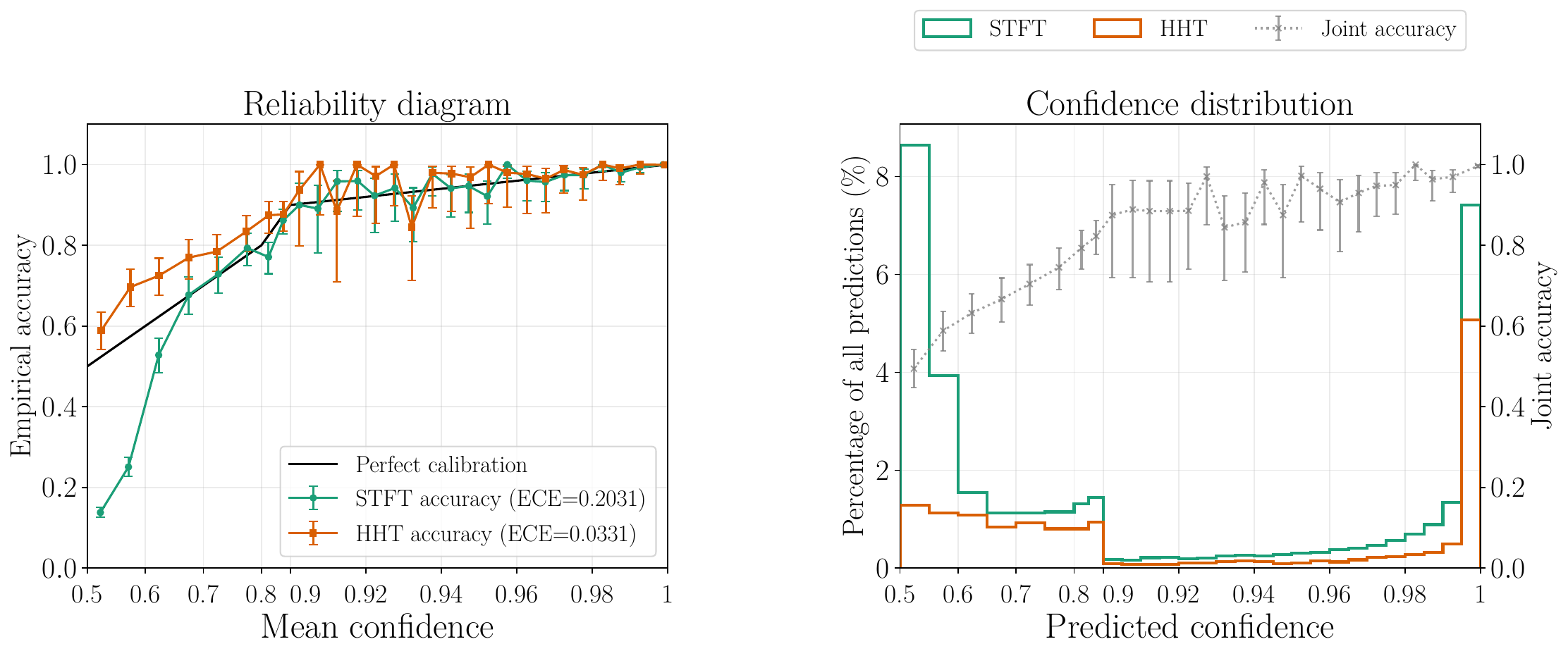}
        \caption{Reconstruction unavailable (33,700 samples)}
        \label{fig:ece-cwb-unavailable}
    \end{subfigure}

    \caption{Reliability diagrams (left) and confidence distributions (right) for the STFT- and HHT-based classifiers evaluated on test signals across all distances, separated according to whether reconstruction is (a) available or (b) unavailable. The reliability diagram compares the mean predicted confidence with the empirical accuracy in ten equal-width confidence bins. A perfectly calibrated model follows the black line, while the Expected Calibration Error (ECE) for each method is reported in the legend. The gray dotted curves show the joint accuracy of the STFT and HHT classifiers. The confidence distributions show the percentage of samples as a function of the predicted confidence. Error bars indicate the 95\% confidence intervals (CIs) for each bin.
    }
    \label{fig:ece}
\end{figure*}

The HHT-based model exhibits a similar overall trend to the STFT-based model, although its classification accuracy is slightly lower. This difference may partly arise from variability introduced by the empirical decomposition in CEEMD. Nevertheless, when predictions are selected according to the confidence derived from the softmax outputs of each classifier, the parenthesized entries in Table~\ref{tab:test_accuracy_by_distance} exceed 99\% at every distance.
To further examine the relationship between prediction confidence and classification accuracy, we evaluate the ECE defined in Eq.~(\ref{eq:ece}).
Figure~\ref{fig:ece}(\subref{fig:ece-cwb-available}) shows the reliability diagrams and confidence distributions for both classifiers. The STFT classifier exhibits a tendency toward overconfident predictions in the confidence range of approximately 0.6--0.9, where the empirical accuracy falls below the predicted confidence. In contrast, this tendency is less pronounced for the HHT classifier, whose empirical accuracy more closely follows the perfect-calibration line.
However, the majority of predictions from both classifiers are concentrated near a confidence of 1. For high-confidence predictions, the joint accuracy of the STFT- and HHT-based classifiers reaches approximately 99\%, indicating that the two distinct analysis approaches provide highly consistent classification results in this regime.

%This result suggests that the HHT representation provides feature representations that are more consistent with the underlying class labels, allowing the classifier to assign confidence scores that better reflect the true probability of correct predictions.

\subsection{Classification without reconstruction}\label{subsec:results_raw}
\begin{figure}[htb!]
        \centering
        \includegraphics[width=\linewidth]{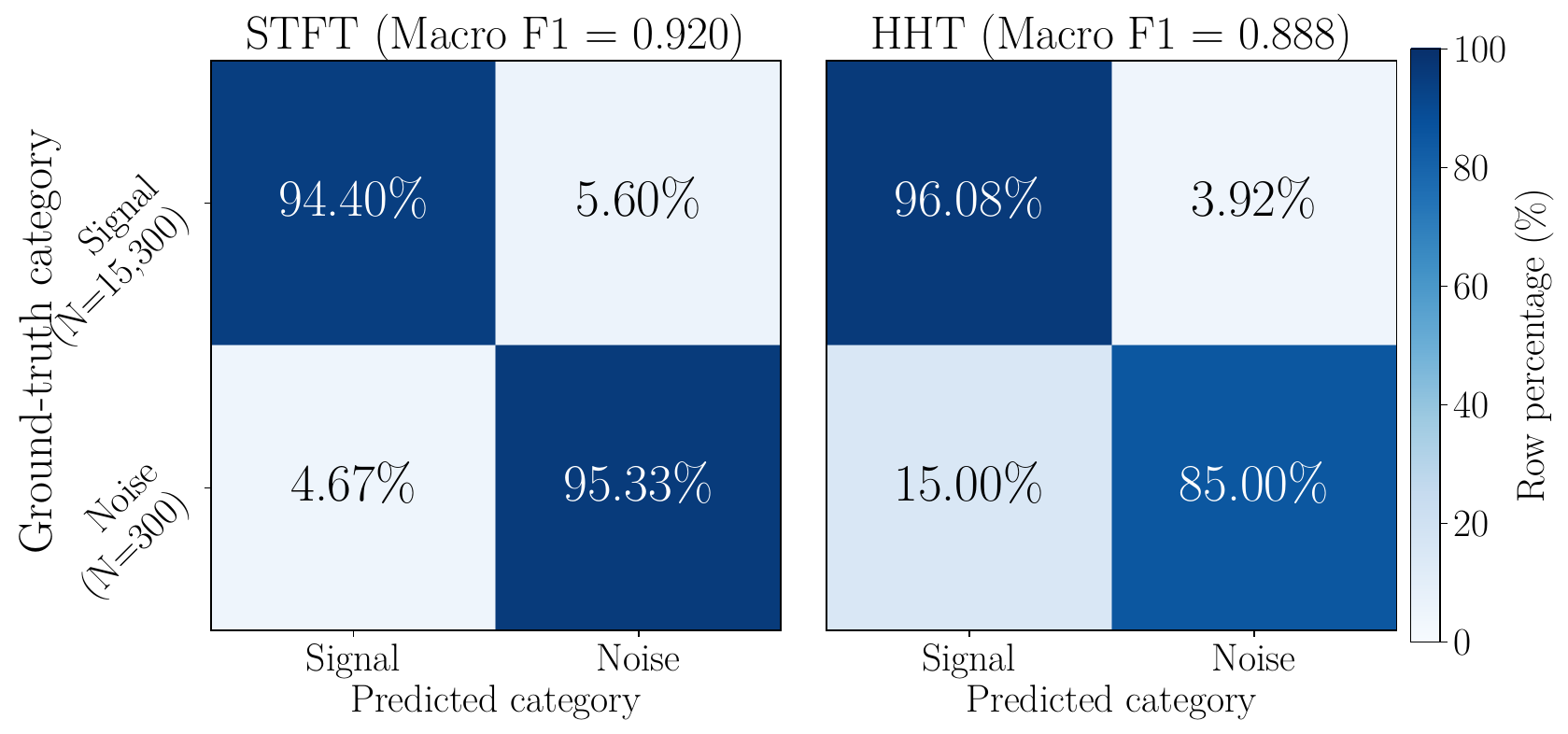}
    \caption{Row-normalized confusion matrices for signal--noise discrimination within $20 \mathrm{kpc}$ using the STFT- and HHT-based raw-input classifiers. The 52-class labels are mapped to two categories: the 51 signal classes are grouped as Signal, and the noise class is retained as Noise.
    All predictions within each distance range are included, without a confidence threshold or a cWB reconstruction. The macro-$F_1$ values evaluate the original 52-class task, not the collapsed binary task.}
    \label{fig:noise_vs_signal_confusion}
\end{figure}

In this section, we evaluate the classification performance of the STFT- and HHT-based models without requiring cWB reconstruction. This approach also allows a noise-only class to be included to account for events without CCSN signals.
A similar trend is observed for both the STFT- and HHT-based classifiers.
As shown in Fig.~\ref{fig:dataset}, cWB reconstructions are unavailable for a fraction of the events. In this analysis, these events are also included in the training and test datasets. Figure~\ref{fig:noise_vs_signal_confusion} shows the confusion matrices for signal--noise discrimination, focusing on events within 20~kpc as representative of Galactic CCSNe. The binary matrices do not assess waveform-class accuracy. The macro-$F_1$ values instead average Eq.~\eqref{eq:f1} over all 52 classes with equal weight. Within 20~kpc, most signal events are correctly identified as signals, with a similar trend observed for both the STFT- and HHT-based classifiers.

\begin{figure}[htb]
    \centering
    \includegraphics[width=\columnwidth]{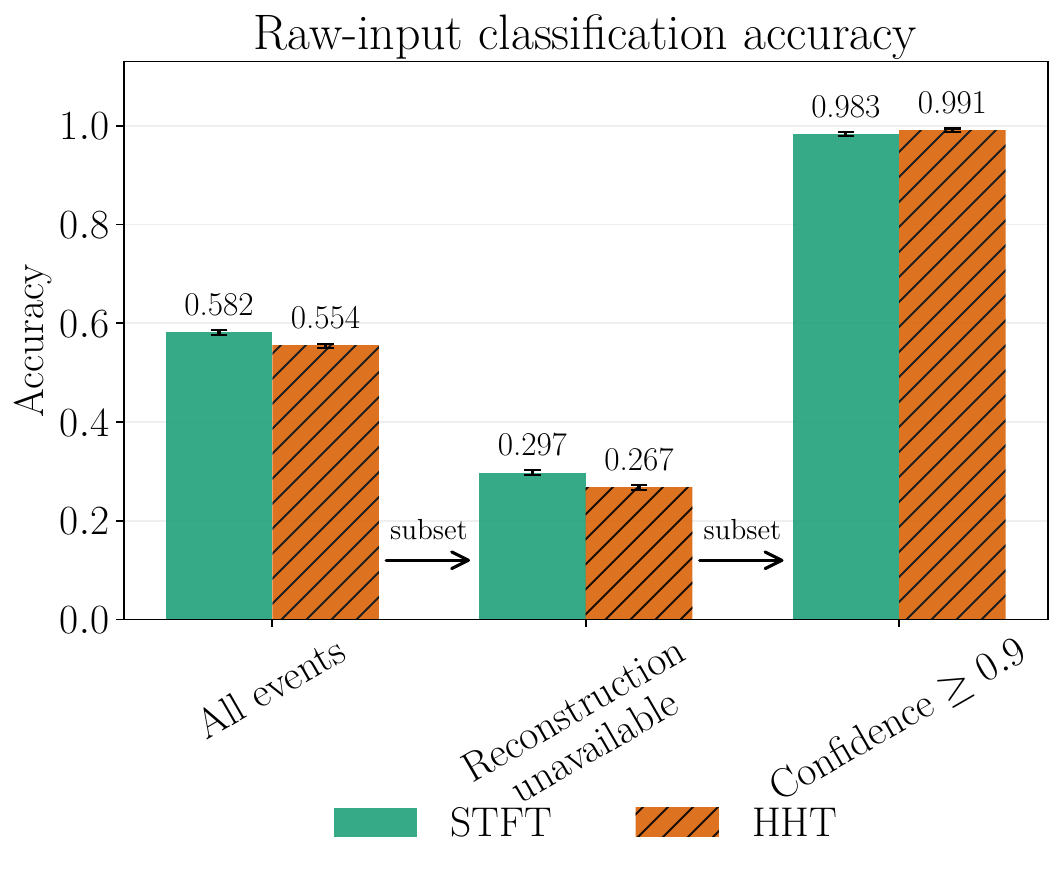}
    \caption{Classification accuracy of the STFT- and HHT-based models evaluated on the 52-class task using the raw-input data. The three groups from left to right represent progressively restricted subsets of the test data: all test events, events for which a cWB reconstruction is unavailable, and high-confidence events ($\geq 0.9$) within the reconstruction-unavailable subset. Error bars show 95\% percentile intervals from 500 bootstrap resamples of the evaluated events.
    }
    \label{fig:stft_hht_accuracy_by_cwb_status}
\end{figure}

Figure~\ref{fig:stft_hht_accuracy_by_cwb_status} further separates the raw-input results according to reconstruction availability. The first pair of bars shows full-sample accuracies below 60\% for both STFT and HHT, where approximately half of the dataset consists of events with available cWB reconstructions.
For cWB reconstruction-unavailable events, accuracies decrease to below 30\%, as shown by the middle pair.
However, selecting high-confidence predictions within this subset increases the accuracy to approximately 99\% for both, as shown by the rightmost pair.
Overall, 15.0\% and 8.2\% of the entire test set, including both reconstruction-available and -unavailable events, are retained as high-confidence predictions for STFT and HHT, respectively.
The reliability diagram for the reconstruction-unavailable subset is shown in Fig.~\ref{fig:ece}(\subref{fig:ece-cwb-unavailable}). For high-confidence predictions, a trend similar to that for reconstruction-available events in Fig.~\ref{fig:ece}(\subref{fig:ece-cwb-available}) is observed.

\begin{figure*}[htb!]
    \centering
    \includegraphics[width=2.0\columnwidth]{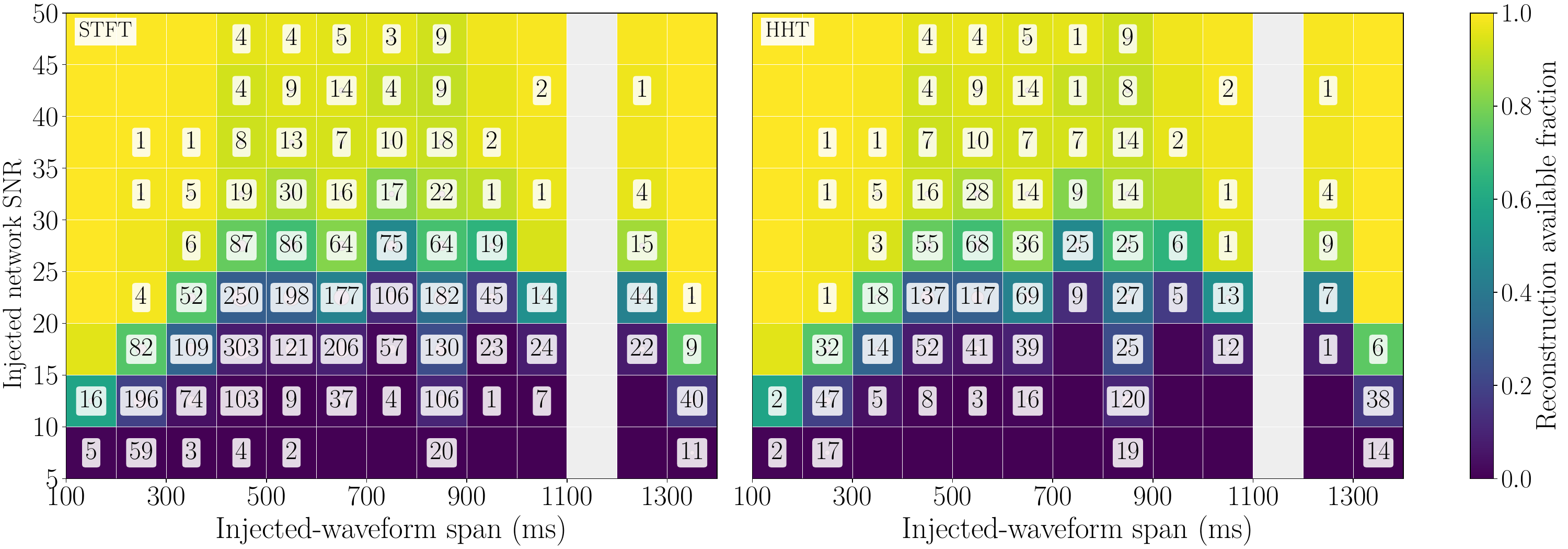}
    \caption{
    High-confidence classification of signals without an reconstruction. The background color shows the reconstruction-availability fraction in bins of injected network SNR and injected-signal duration, calculated from the full dataset across all distances. Overlaid markers indicate test events with unavailable reconstructions that were correctly classified by the raw-input STFT and HHT models with confidence \(\geq0.9\). Marker area increases with event count, and numbers give the count in each bin.
    }
    \label{fig:test_unavailable_conf_gt09_correct_snr_duration}
\end{figure*}

Figure~\ref{fig:test_unavailable_conf_gt09_correct_snr_duration} shows the number of high-confidence predictions that are correctly classified, corresponding to the data in Fig.~\ref{fig:ece}(\subref{fig:ece-cwb-unavailable}), as a function of the injected network SNR and waveform span. Even at the same injected network SNR, the reconstructed SNR can decrease depending on the signal duration and time--frequency complexity~\cite{szczepanczyk2021detecting}. The correctly classified events are concentrated primarily near the transition region where the reconstruction fraction begins to decrease. This suggests that independently applying different analysis methods can provide complementary information.

\subsection{Joint correctness of the two classifiers}\label{subsec:joint}
Table~\ref{tab:paired_joint_accuracy} reports STFT and HHT predictions on matched events, using the event identifier, 
as defined in Eq.~(\ref{eq:joint_accuracy}), for common test events. This comparison evaluates the consistency of the two classification approaches applied to the same underlying GW events. For high-confidence predictions, both raw and cWB inputs exhibit high joint accuracy, indicating that the STFT- and HHT-based classifiers consistently classify the same events correctly despite the different input processing.

\begin{table}[htb]
    \centering
    \caption{Joint classification performance for paired events. The raw and cWB samples originate from the same test events, with the cWB sample restricted to events with an available reconstruction. The same events are used for the STFT and HHT representations. Conf. denotes events for which both classifiers have confidence $\geq 0.9$, and Cov. denotes the fraction of paired events retained within each input setting.}
    \label{tab:paired_joint_accuracy}
    \begin{tabular}{llccc}
        \hline
        Input & Subset & Samples (Cov.) & Joint acc.& STFT / HHT acc.\\
        \hline
        Raw & All              & 57,200 & 51.99\% & 58.16\% /55.43\% \\
        Raw & Conf. & 23,251 (40.6\%)  & 99.75\% & 99.97\% /99.77\% \\
        \addlinespace[4pt]
        cWB & All              & 23,498 & 94.37\% & 97.81\% /95.04\% \\
        cWB & Conf. & 20,640 (87.8\%) & 99.77\% & 99.40\% /99.70\% \\
        \hline
    \end{tabular}
\end{table}

\section{Discussion}\label{sec6}
The results show that both representations retain sufficient information to distinguish among many numerical CCSN waveforms after cWB reconstruction. The STFT-based classifier provides higher and more stable performance than the HHT-based classifier over most of the distance range.
For HHT, CEEMD may be sensitive to subtle differences among waveforms, making it challenging to achieve consistently effective decomposition using the same CEEMD parameter settings.
Nevertheless, the neural network appears to compensate for some of these variations in the resulting representations.
In particular, the HHT-based classifier has a smaller ECE than the STFT-based classifier on the evaluated samples.
Furthermore, the high joint accuracy shown in Table~\ref{tab:paired_joint_accuracy} indicates that the two classifiers consistently identify many of the same events correctly.
These results suggest that applying multiple analysis pipelines could provide a useful complementary assessment in future observations.

The reconstructed and raw-input analyzes provide complementary views of the classification performance.
A reconstruction suppresses noise and can make the waveform structure more accessible to the classifier, as illustrated by the waveform and spectrogram examples in Figs.~\ref{fig:mezzacappa23_d15_3d} and \ref{fig:m39_STFT}. The UMAP visualizations in Appendix~\ref{appendix:umap} further provide a qualitative view of the effectiveness of the cWB reconstruction and the subsequent time--frequency representations in separating waveform classes.
For future observations, it will be important to compare the classification results obtained with different analysis pipelines, including more recent developments such as cWB-XP~\cite{2022APS..APRE17004K}, to assess the consistency and robustness of the inferred waveform classification.

The CCSN rate within the Milky Way is estimated to be
$1.63\pm0.46\,(100\,{\rm yr})^{-1}$~\cite{Rozwadowska:2020nab}.
For comparison, the total supernova rate, including both core-collapse
and Type Ia supernovae, in the Large and Small Magellanic Clouds
combined is estimated to be $0.25$--$0.46\,(100\,{\rm yr})^{-1}$
~\cite{2010MNRAS.407.1314M}. Although including the Magellanic Clouds is expected to increase the CCSN event rate only modestly, CCSN events with sufficiently high network SNRs are expected to be detectable at these distances.

\begin{figure*}[htb!]
    \centering
    \begin{subfigure}{\columnwidth}
        \centering
        \includegraphics[width=\linewidth]{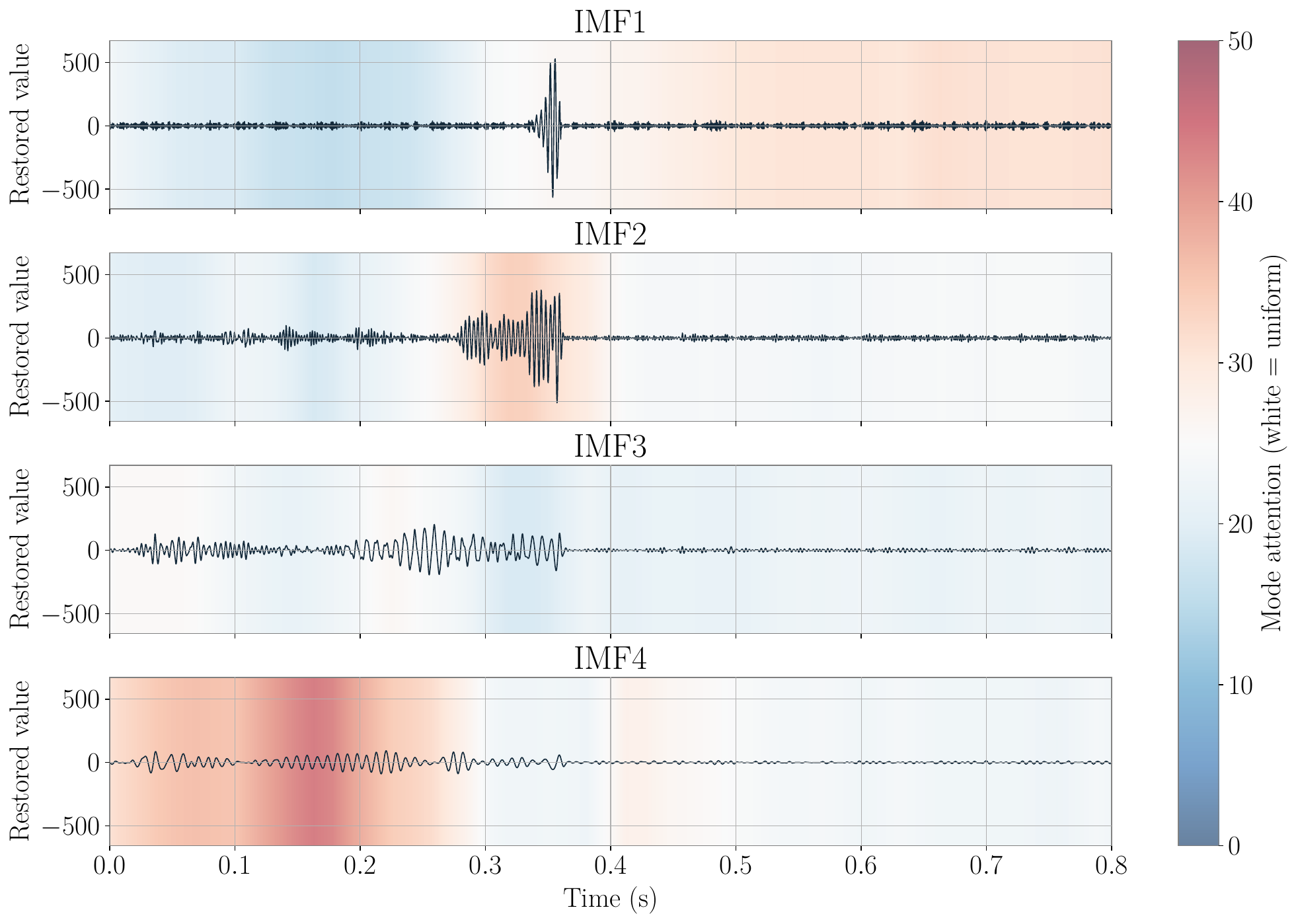}
        \caption{IMF mode attention on time-series}
        \label{fig:attn_imf}
    \end{subfigure}
    \hfill
    \begin{subfigure}{\columnwidth}
        \centering
        \includegraphics[width=\linewidth]{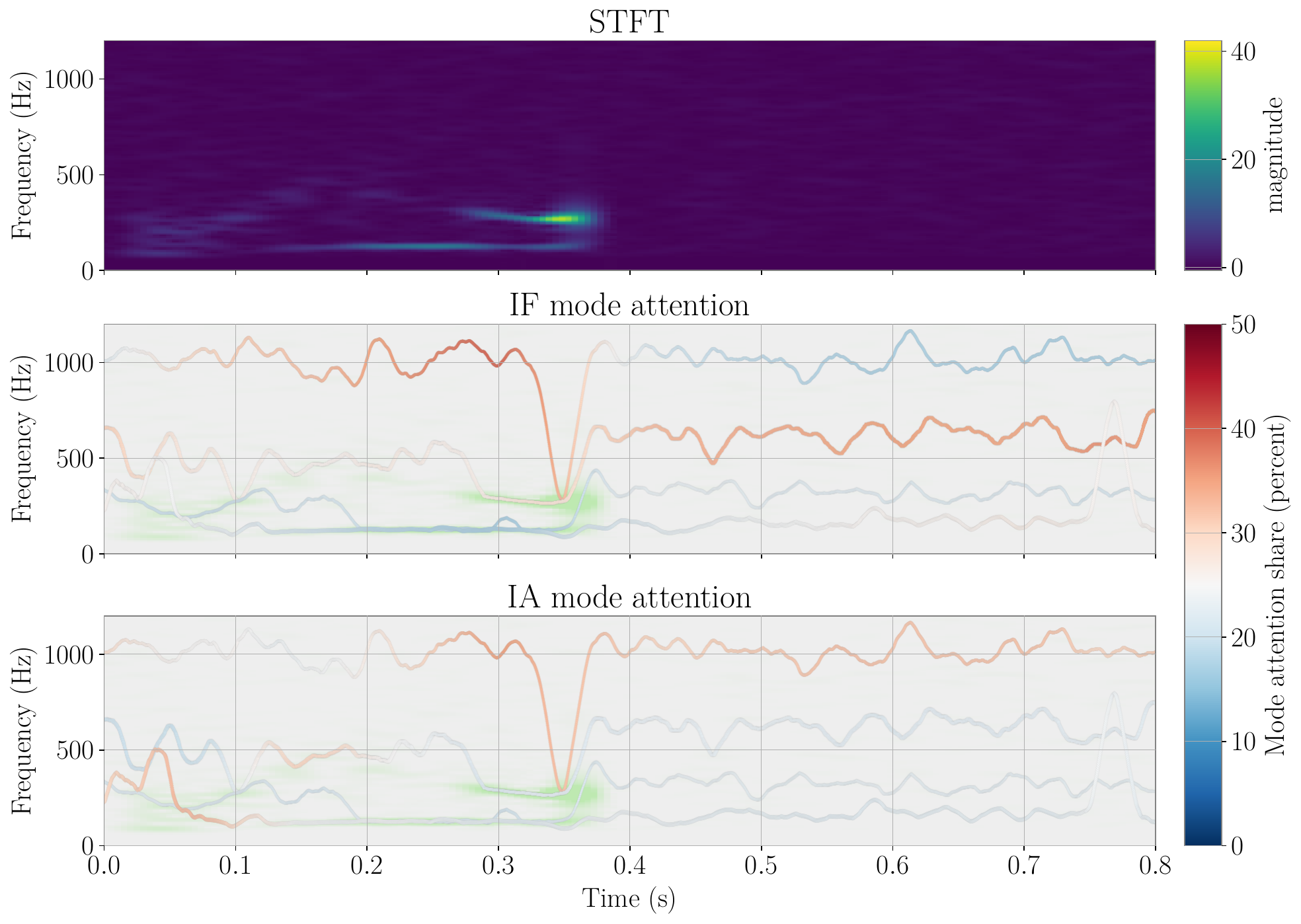}
        \caption{IF and IA mode attention in the time--frequency plane}
        \label{fig:attn_ia_if}
    \end{subfigure}
    \caption{Attention maps for the Kuroda et al.\ (2016) \texttt{SFHx} waveform in ET1 at 10~kpc. (a) IMF attention. (b) STFT as a reference (top), IF attention (middle) and IA attention (bottom), displayed along the instantaneous-frequency tracks of the corresponding IMFs. Colors indicate the attention received by each mode, averaged over attention heads, query modes, and axial attention layers. At each time, the shares sum to 100\% across the four modes; 25\% corresponds to uniform attention. Patch-level weights are linearly interpolated onto the waveform time grid before averaging across layers. The green background in panel (b) shows the STFT of the same event as a reference.}
    \label{fig:attention_m39}
\end{figure*}

\begin{figure*}[htb!]
    \centering
    \begin{subfigure}{\columnwidth}
        \centering
        \includegraphics[width=\linewidth]{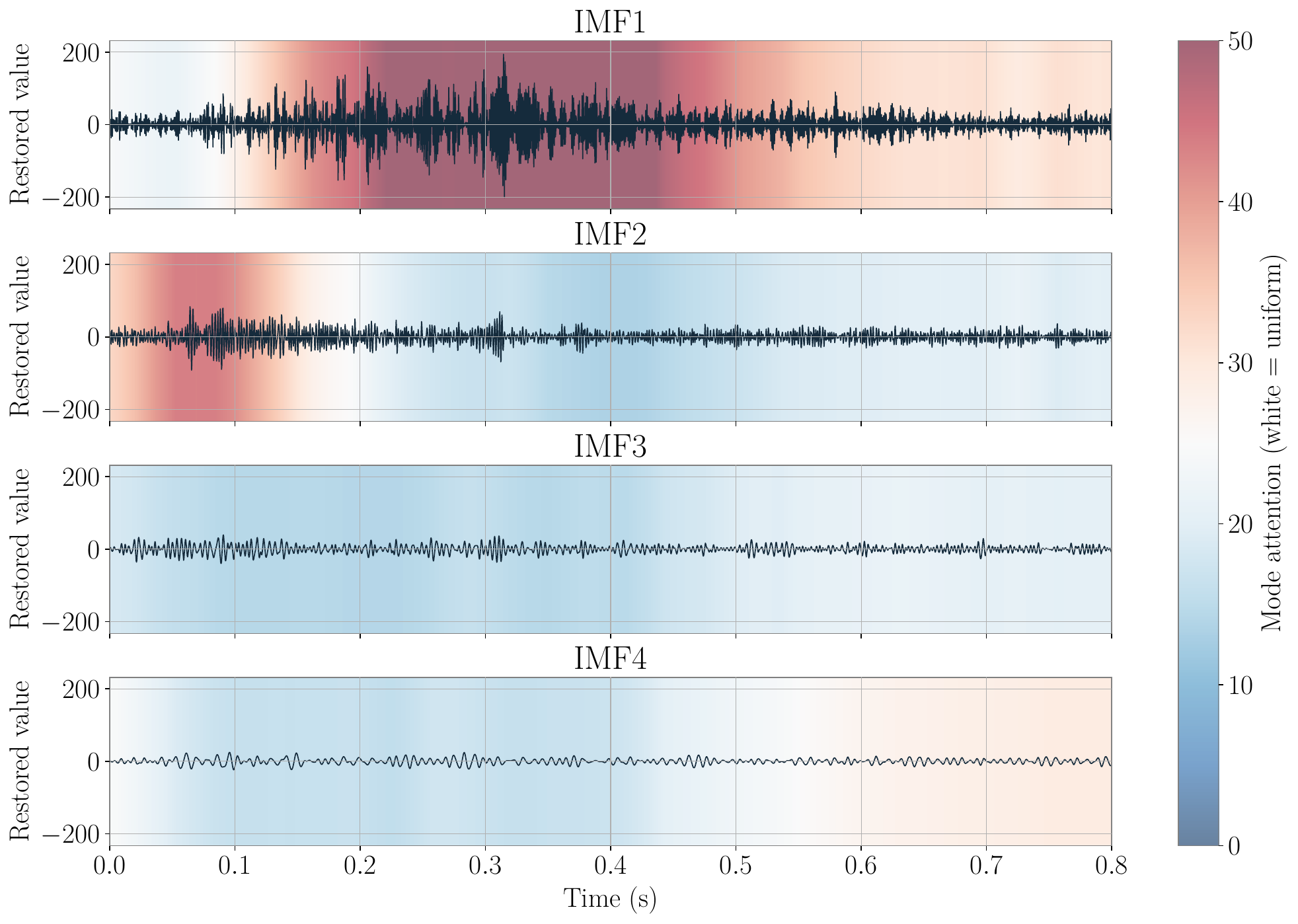}
        \caption{IMF mode attention on time-series}
        \label{fig:attn_imf_s35}
    \end{subfigure}
    \hfill
    \begin{subfigure}{\columnwidth}
        \centering
        \includegraphics[width=\linewidth]{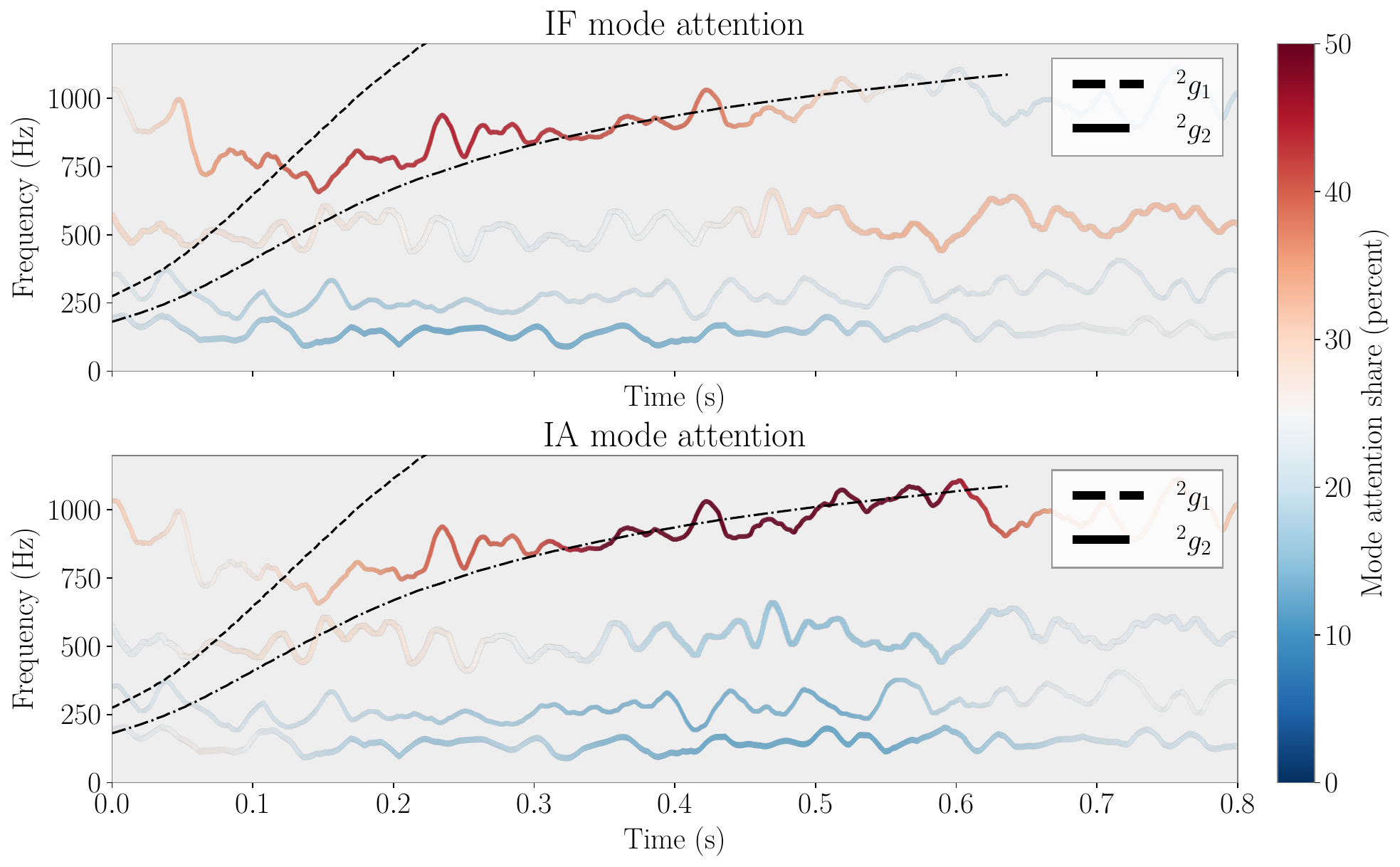}
        \caption{IF and IA mode attention in the time--frequency plane}
        \label{fig:attn_ia_if_s35}
    \end{subfigure}
    \caption{
    Attention maps for the Powell et al. (2018) \texttt{s3.5\_pns} waveform at 10 kpc in ET1. (a) IMF time series overlaid on their respective mode-attention weights. (b) IF (top) and IA (bottom) mode-attention weights displayed along the IF tracks. The normalization and averaging follow Fig.~\ref{fig:attention_m39}. The black dashed and dash-dotted curves in panel (b) show the reference frequencies of the $^{2}g_{1}$ and $^{2}g_{2}$ modes calculated from the simulation's PNS properties using universal relations.}
    \label{fig:attention_s35}
\end{figure*}

Previous analyses using CNNs and class activation maps have shown that neural networks tend to focus on regions with strong amplitudes in STFT representations~\cite{2023PhRvD.108l3033S}.
Similarly, the attention weights defined in Eq.~(\ref{eq:attention_weights}) visualize the regions emphasized by the model during classification.
Figure~\ref{fig:attention_m39} visualizes how the cWB-input HHT transformer allocates mode attention for the waveform from Kuroda et al.(2016)~\cite{Kuroda2016}. The SASI mode appears stationary around 100~Hz, and its characteristic time--frequency structure has been identified by HHT in a previous study~\cite{Takeda2021}. The attention map shows a strong concentration on IMF4, suggesting that the Transformer emphasizes features associated with the SASI.
%In contrast, the attention in the IF and IA representations appears to be distributed relatively uniformly over the entire region.
Furthermore, Fig.~\ref{fig:attention_s35} shows the attention map with the universal relation associated with the physical properties of the proto-neutron star~\cite{Torres2021}.
The attention is predominantly concentrated on the IF of IMF1. However, the attention across the IMFs exhibits a temporal evolution, shifting from the lower-frequency IMF4 toward the higher-frequency IMF1 as time progresses. The attention tracks the time--frequency evolution associated with the $g$-mode.
Around $t=0$~s, the attention is distributed relatively uniformly over the time--frequency map, whereas at later times it becomes increasingly concentrated on the IF and IA of IMF1, which are associated with the $^2g_2$ mode. 
Because features such as the $g$-mode drift across different regions of the time--frequency domain, combining attention maps with the IMFs may help identify candidate waveform components associated with these features.
As a future direction, the features highlighted by the attention mechanism could provide a basis for extracting information on physical parameters associated with the CCSN explosion mechanism from the analyzed GW signals.

Ultimately, establishing consistency across different analysis pipelines, including those based on neural networks, will be essential for advancing our understanding of the CCSN explosion mechanism.

\section{Conclusion}\label{sec7}
We have investigated deep-learning recognition of 51 waveforms from 3D CCSN simulations projected onto a triangular ET detector configuration. The two models, based on STFT spectrograms with ConvNeXt V2 and HHT components with an axial attention transformer, respectively, provide consistent classification results despite their distinct time--frequency representations and neural network architectures.
Both distinguish waveforms in the catalog with high accuracy, demonstrating the effectiveness of deep learning for discriminating among CCSN waveforms.

The present study focuses on waveform-level classification rather than direct inference of physical parameters. Since numerical CCSN waveforms are generated using different hydrodynamic solvers, microphysics prescriptions, progenitor models, and numerical resolutions, successful classification does not necessarily imply that the network has learned disentangled physical parameters. Future work should investigate hierarchical labeling schemes based on physical properties such as EOS, rotation, magnetic field strength, explosion outcome, and SASI activity, to determine which physical features are robustly encoded in the learned representations.

A key direction for future work is to improve generalization to unseen waveforms and to connect the features extracted from GW signals with the physical properties predicted by theoretical models and numerical simulations. Extending the present classification framework toward the inference of physical parameters will be important for extracting further information on CCSN dynamics and for advancing our understanding of CCSN physics.

\appendix
\section{HHT Spectrum}\label{appendix:hht}
\begin{figure}[htb]
    \centering
    \includegraphics[width=\columnwidth]{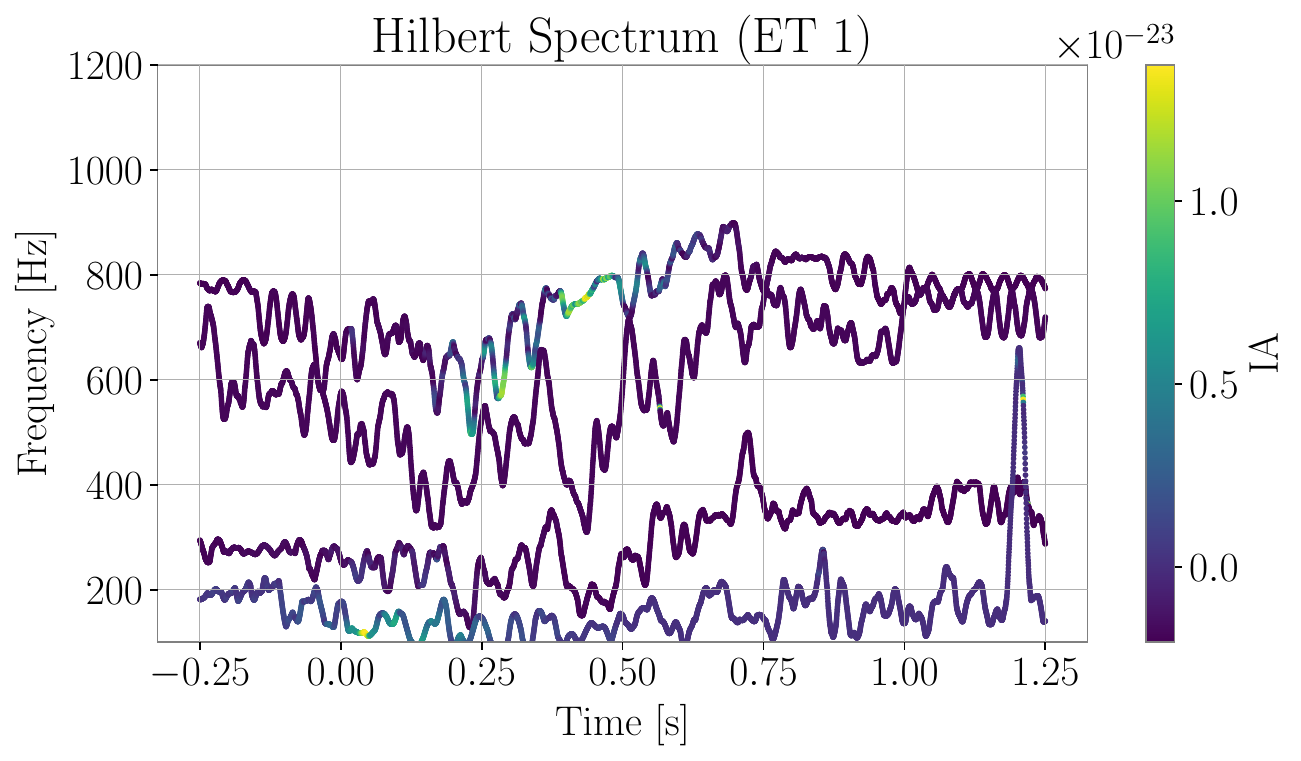}
    \caption{A visualization of the HHT spectrum for the \texttt{m39} model at a source distance of 50~kpc.
    }
    \label{fig:m39_HHT_map}
\end{figure}

Figure~\ref{fig:m39_HHT_map} illustrates the HHT spectrum of the \texttt{m39} model. The HHT spectrum captures time--frequency evolution similar to that shown by the STFT spectrogram in Fig.~\ref{fig:m39_STFT}, particularly near 800~Hz. Note that this figure is included only to visualize the HHT representation. Our HHT-based classifier does not use this image directly; instead, it uses the time-series IMFs, IAs, and IFs as input features, which are the essential components of the HHT.

\section{UMAP clustering with and without cWB pipeline}\label{appendix:umap}
\begin{figure*}[t]
    \centering

    \begin{subfigure}{\columnwidth}
        \centering
        \includegraphics[width=\linewidth]{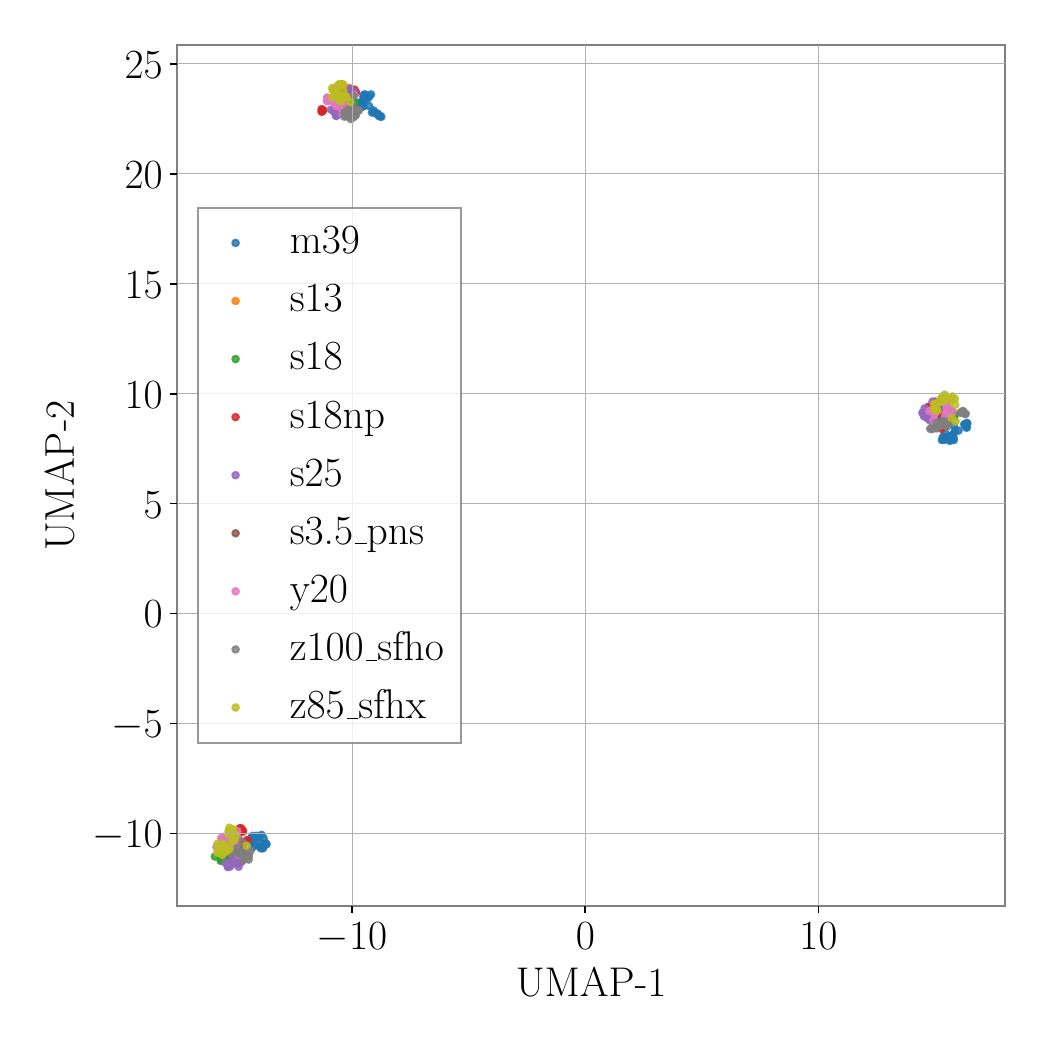}
        \caption{Original waveform (ET1)}
        \label{fig:umap_et}
    \end{subfigure}
    \hfill
    \begin{subfigure}{\columnwidth}
        \centering
        \includegraphics[width=\linewidth]{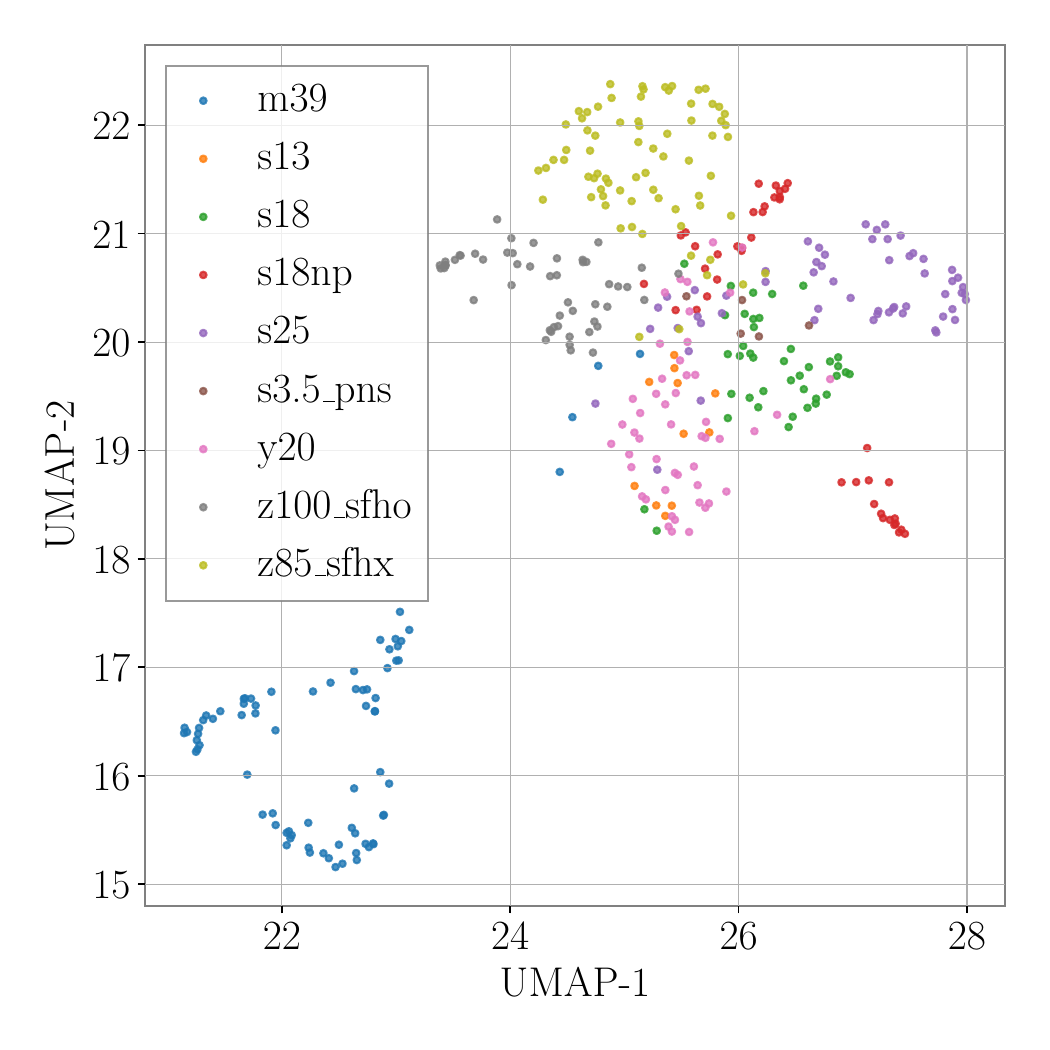}
        \caption{cWB}
        \label{fig:umap_cwb}
    \end{subfigure}

    \vspace{0.5em}

    \begin{subfigure}{\columnwidth}
        \centering
        \includegraphics[width=\linewidth]{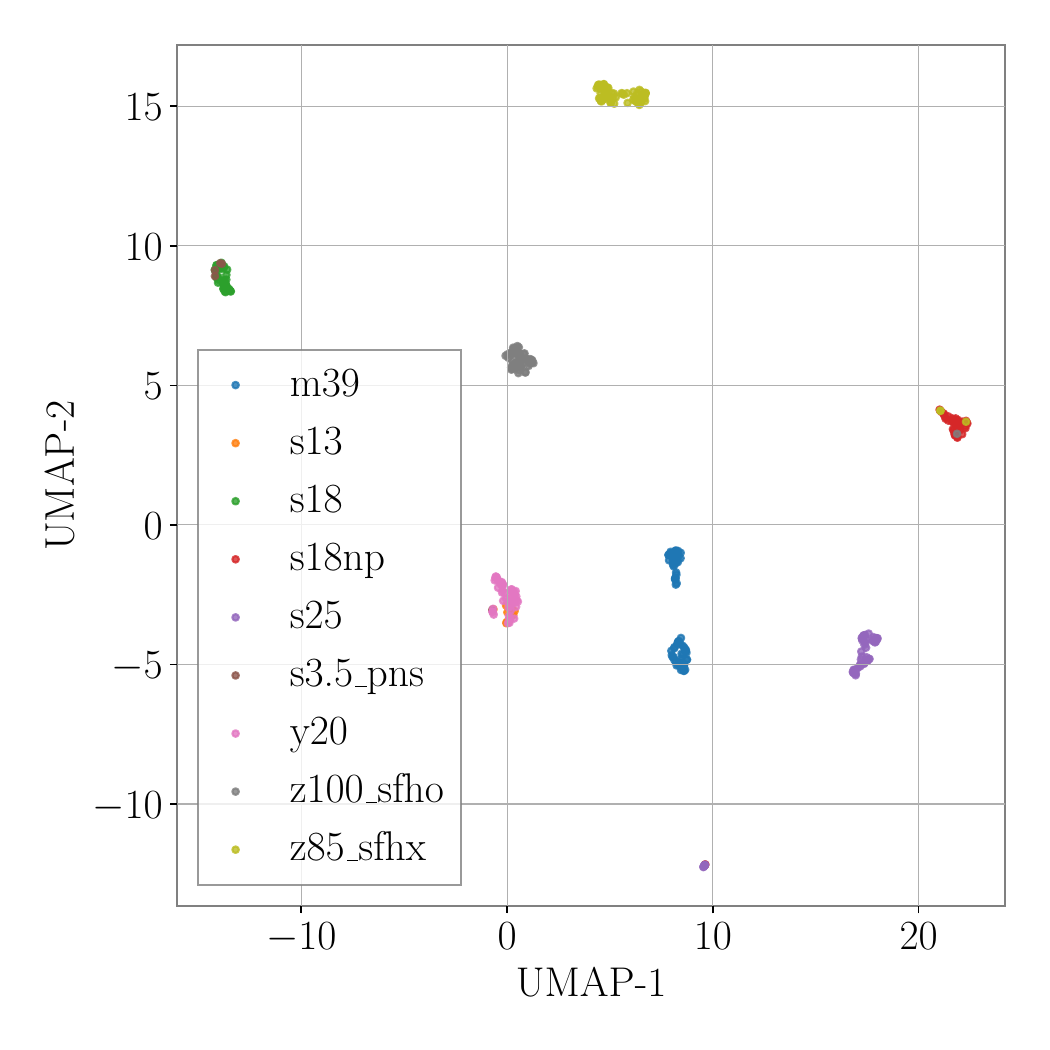}
        \caption{cWB + STFT}
        \label{fig:umap_stft}
    \end{subfigure}
    \hfill
    \begin{subfigure}{\columnwidth}
        \centering
        \includegraphics[width=\linewidth]{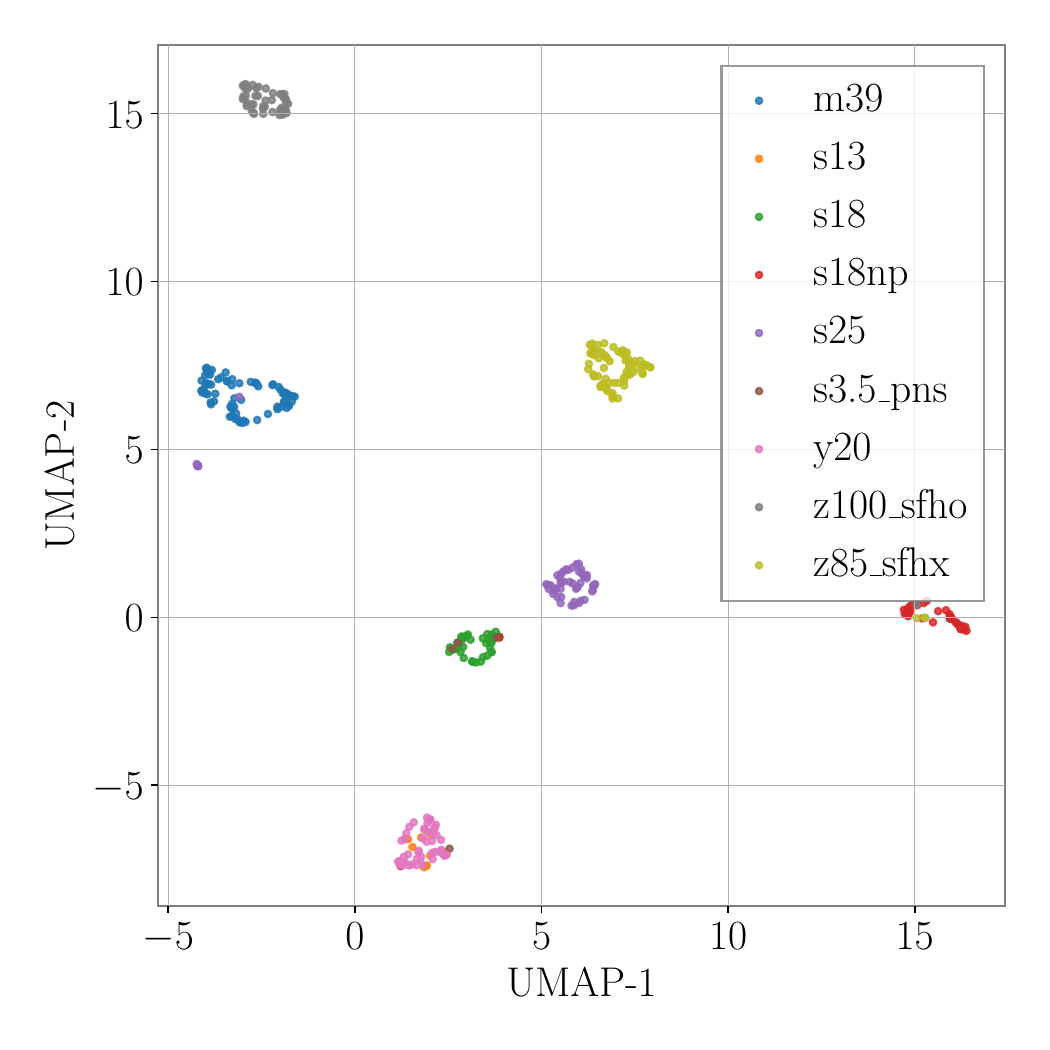}
        \caption{cWB + HHT}
        \label{fig:umap_hht}
    \end{subfigure}

    \caption{
    UMAP visualization of the waveform representations for nine representative CCSN waveform classes at a source distance of 50~kpc.
    (a) Original waveforms, (b) cWB-reconstructed waveforms, (c) STFT representations of the cWB-reconstructed waveforms, and (d) HHT representations of the cWB-reconstructed waveforms by concatenating all IMF, IF, and IA timeseries.
    }
    \label{fig:umap}
\end{figure*}

Uniform Manifold Approximation and Projection for Dimension Reduction (UMAP)~\cite{2018arXiv180203426M} is a nonlinear dimensionality reduction algorithm that preserves both the local neighborhood structure and the global topology of high-dimensional data in a low-dimensional embedding.
In this study, UMAP is used to visualize and compare the distributions of different waveform representations, including (a) the original waveforms, (b) the cWB-reconstructed waveforms, and the time-frequency representations of the cWB-reconstructed waveforms obtained using (c) STFT and (d) HHT, as shown in Fig.~\ref{fig:umap}. For simplicity, only nine representative CCSN waveform classes are used here.
The original waveforms exhibit poor separation between classes in the embedding space.
After the cWB, distinct clusters begin to emerge, indicating that the reconstruction process enhances class-discriminative features.
The separation becomes substantially clearer for the STFT- and HHT-based time-frequency representations, where most waveform classes form well-defined clusters.
These results qualitatively demonstrate that each stage of the proposed analysis pipeline progressively improves the discriminability of the waveform representations, thereby providing more informative inputs for the subsequent classification network.

\section{ConvNext V2 architecture}\label{appendix:cnn}
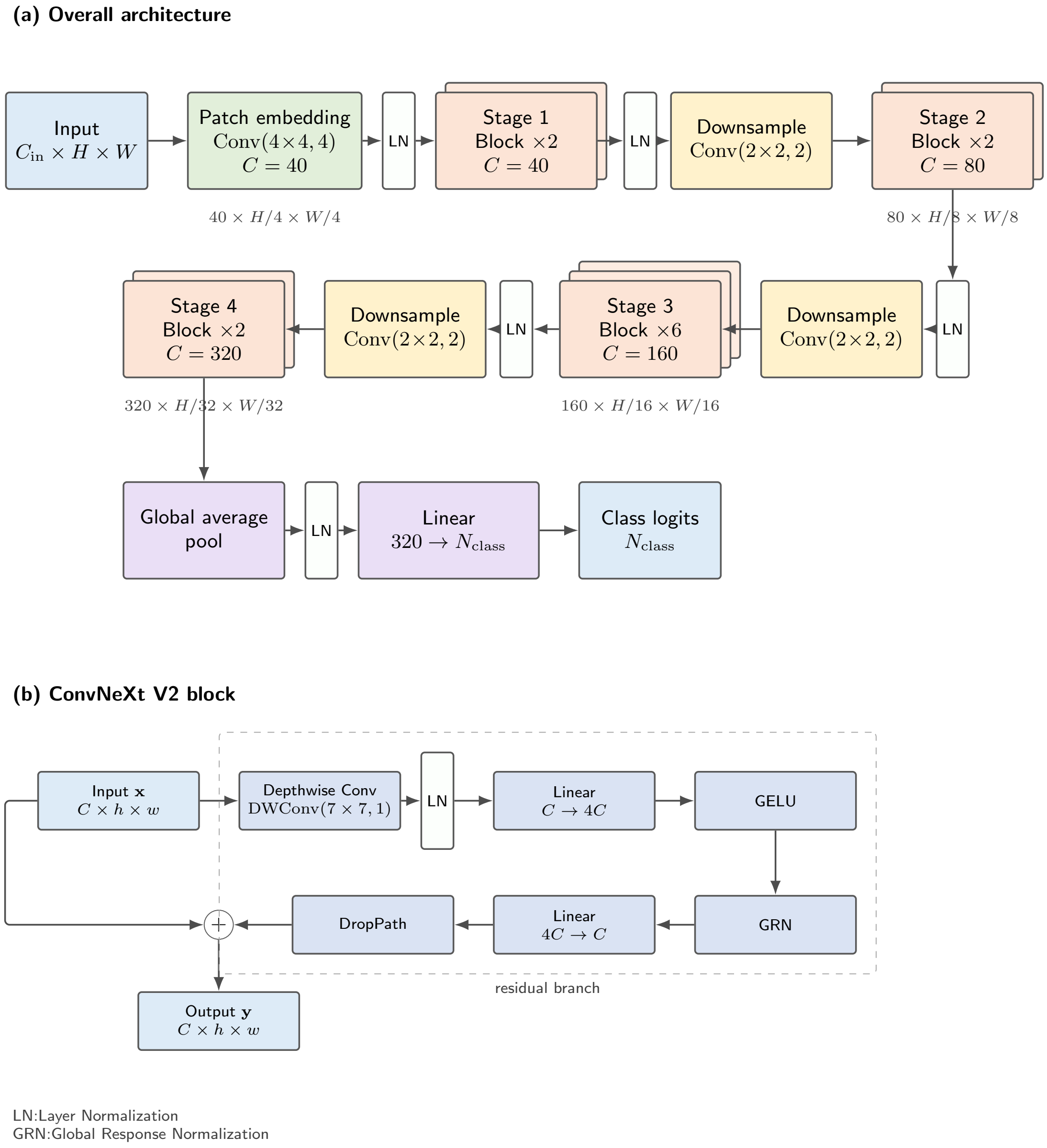
\begin{figure*}[tb!]
\centering
\resizebox{0.85\textwidth}{!}{% TikZ fragment for use with \input{convnextv2_architecture}.
% Required in the main document preamble:
% \usepackage{tikz}
% \usetikzlibrary{arrows.meta,backgrounds,calc,fit,positioning}
\definecolor{inputblue}{RGB}{221,235,247}
\definecolor{embedgreen}{RGB}{226,239,218}
\definecolor{stageorange}{RGB}{252,228,214}
\definecolor{downyellow}{RGB}{255,242,204}
\definecolor{headpurple}{RGB}{234,223,246}
\definecolor{blockblue}{RGB}{218,227,243}
\definecolor{axpatch}{RGB}{226,239,218}

\tikzset{
  flow/.style={-{Latex[length=2.2mm]}, line width=0.75pt, draw=black!70},
  box/.style={draw=black!65, rounded corners=1.5pt, line width=0.75pt,
    align=center, minimum height=15mm, font=\sffamily\small},
  op/.style={box, minimum width=25mm, minimum height=9mm, font=\sffamily\scriptsize},
  lnbox/.style={box, fill=green!1, minimum width=5mm,
    minimum height=15mm, inner xsep=1pt, font=\sffamily\scriptsize},
  skip/.style={flow, rounded corners=2pt},
  note/.style={font=\sffamily\scriptsize, text=black!75, align=center}
}

\begin{tikzpicture}[node distance=5mm and 6mm]
  % Panel (a): complete network
  \node[font=\sffamily\bfseries, anchor=west] (atitle) {(a) Overall architecture};

  \node[box, fill=inputblue, minimum width=22mm, anchor=north west]
    (input) at ($(atitle.south west)+(0,-9mm)$) {Input\\$C_{\mathrm{in}}\times H\times W$};
  \node[box, fill=embedgreen, minimum width=27mm, right=of input]
    (stem) {Patch embedding\\$\mathrm{Conv}(4\!\times\!4,4)$\\$C=40$};
  \node[lnbox, right=3mm of stem] (stemln) {LN};
  \node[box, fill=stageorange, minimum width=25mm, right=3mm of stemln]
    (s1) {Stage 1\\Block $\times 2$\\$C=40$};
  \node[lnbox, right=4mm of s1] (d1ln) {LN};
  \node[box, fill=downyellow, minimum width=25mm, right=2mm of d1ln]
    (d1) {Downsample\\ $\mathrm{Conv}(2\!\times\!2,2)$};
  \node[box, fill=stageorange, minimum width=25mm, right=of d1]
    (s2) {Stage 2\\Block $\times 2$\\$C=80$};

  \node[lnbox, below=14mm of s2] (d2ln) {LN};
  \node[box, fill=downyellow, minimum width=25mm, left=2mm of d2ln]
    (d2) {Downsample\\ $\mathrm{Conv}(2\!\times\!2,2)$};
  \node[box, fill=stageorange, minimum width=25mm, left=of d2]
    (s3) {Stage 3\\Block $\times 6$\\$C=160$};
  \node[lnbox, left=4mm of s3] (d3ln) {LN};
  \node[box, fill=downyellow, minimum width=25mm, left=2mm of d3ln]
    (d3) {Downsample\\ $\mathrm{Conv}(2\!\times\!2,2)$};
  \node[box, fill=stageorange, minimum width=25mm, left=of d3]
    (s4) {Stage 4\\Block $\times 2$\\$C=320$};
  \node[box, fill=headpurple, minimum width=25mm, below=16mm of s4]
    (pool) {Global average\\pool};
  \node[lnbox, right=3mm of pool] (headln) {LN};
  \node[box, fill=headpurple, minimum width=28mm, right=3mm of headln]
    (head) {Linear\\$320\to N_{\mathrm{class}}$};
  \node[box, fill=inputblue, minimum width=22mm, right=of head]
    (output) {Class logits\\$N_{\mathrm{class}}$};

  % Offset copies visualize the repeated blocks in each stage.
  \begin{scope}[on background layer]
    \node[box, fill=stageorange!82, minimum width=25mm]
      at ($(s1.center)+(1.5mm,1.5mm)$) {};
    \node[box, fill=stageorange!82, minimum width=25mm]
      at ($(s2.center)+(1.5mm,1.5mm)$) {};
    \node[box, fill=stageorange!72, minimum width=25mm]
      at ($(s3.center)+(3mm,3mm)$) {};
    \node[box, fill=stageorange!82, minimum width=25mm]
      at ($(s3.center)+(1.5mm,1.5mm)$) {};
    \node[box, fill=stageorange!82, minimum width=25mm]
      at ($(s4.center)+(1.5mm,1.5mm)$) {};
  \end{scope}

  \draw[flow] (input) -- (stem);
  \draw[flow] (stem) -- (stemln);
  \draw[flow] (stemln) -- (s1);
  \draw[flow] (s1) -- (d1ln);
  \draw[flow] (d1ln) -- (d1);
  \draw[flow] (d1) -- (s2);
  \draw[flow] (s2.south) -- (d2ln.north);
  \draw[flow] (d2ln) -- (d2);
  \draw[flow] (d2) -- (s3);
  \draw[flow] (s3) -- (d3ln);
  \draw[flow] (d3ln) -- (d3);
  \draw[flow] (d3) -- (s4);
  \draw[flow] (s4.south) -- (pool.north);
  \draw[flow] (pool) -- (headln);
  \draw[flow] (headln) -- (head);
  \draw[flow] (head) -- (output);

  \node[note, below=2mm of stem] {$40\times H/4\times W/4$};
  \node[note, below=2mm of s2] {$80\times H/8\times W/8$};
  \node[note, below=2mm of s3] {$160\times H/16\times W/16$};
  \node[note, below=2mm of s4] {$320\times H/32\times W/32$};

  % Panel (b): one ConvNeXt V2 block
  \coordinate (panelb) at ($(atitle.west |- pool.south)+(0,-18mm)$);
  \node[font=\sffamily\bfseries, anchor=west] (btitle) at (panelb)
    {(b) ConvNeXt V2 block};
  \node[op, fill=inputblue, anchor=north west] (bin)
    at ($(btitle.south west)+(5mm,-9mm)$) {Input $\mathbf{x}$\\$C\times h\times w$};
  \node[op, fill=blockblue, right=of bin]
    (dw) {Depthwise Conv\\$\mathrm{DWConv}(7\times7,1)$};
  \node[lnbox, right=3mm of dw] (ln) {LN};
  \node[op, fill=blockblue, right=of ln] (pw1) {Linear\\$C\to4C$};
  \node[op, fill=blockblue, right=of pw1] (gelu) {GELU};
  \node[op, fill=blockblue, below=10mm of gelu] (grn) {GRN};
  \node[op, fill=blockblue, left=of grn] (pw2) {Linear\\$4C\to C$};
  \node[op, fill=blockblue, left=of pw2] (dp) {DropPath};
  \node[circle, draw=black!65, fill=white, inner sep=1.2pt, left=9mm of dp,
    font=\sffamily\small] (add) {$+$};
  \node[op, fill=inputblue, below=8mm of add]
    (bout) {Output $\mathbf{y}$\\$C\times h\times w$};

  \draw[flow] (bin) -- (dw);
  \draw[flow] (dw) -- (ln);
  \draw[flow] (ln) -- (pw1);
  \draw[flow] (pw1) -- (gelu);
  \draw[flow] (gelu.south) -- (grn.north);
  \draw[flow] (grn) -- (pw2);
  \draw[flow] (pw2) -- (dp);
  \draw[flow] (dp) -- (add);
  \draw[flow] (add.south) -- (bout.north);
  \draw[skip] (bin.west) -- ++(-5mm,0) |- (add.west);

  \node[draw=black!35, dashed, rounded corners=2pt, fit=(dw)(ln)(pw1)(gelu)(grn)(pw2)(dp),
    inner sep=3mm, label={[note]below:residual branch}] {};

  \node[note, anchor=north west, align=left] at ($(btitle.west |- bout.south)+(0,-8mm)$)
    {LN:Layer Normalization\\
     GRN:Global Response Normalization};
\end{tikzpicture}}
\caption{The ConvNeXt V2 architecture for STFT inputs. The model contains $3{,}403{,}771$ trainable parameters for $N_{\mathrm{class}}=51$. 
(a) Overall network architecture. The values shown below each stage indicate the dimensions of the output feature maps. $C_{\mathrm{in}}$ denotes the number of input channels, while $H$ and $W$ denote the height and width of the input image, respectively. $\mathrm{Conv}(K \times K, s)$ denotes a convolutional layer with kernel size $K \times K$ and stride $s$. 
(b) Internal structure of a ConvNeXt V2 block, where $C$ denotes the number of channels in the corresponding stage, and $h$ and $w$ denote the height and width of the feature map in the embedding space, respectively. DropPath stochastically drops the residual branch with a probability of $0.1$.}
\label{fig:convnextv2-architecture}
\end{figure*}

We employed a ConvNeXt V2 classifier comprising a four-stage hierarchical
feature extractor shown in Fig.~\ref{fig:convnextv2-architecture}. The channel dimensions of the four stages were set to
$(40,80,160,320)$, and the corresponding numbers of ConvNeXt V2 blocks were
$(2,2,6,2)$. 

\section{Axial attention Transformer architecture}\label{appendix:axial}
For each temporal window and detector, the IMF, instantaneous frequency (IF), and instantaneous amplitude (IA) are independently projected into token embeddings by a linear layer.

The hierarchical axial attention blocks share their parameters across the IMF, IF, and IA components. Each component is nevertheless processed independently within these blocks; the axial self-attention operations mix information only along the mode, temporal-patch, and detector axes. Information across the three components is integrated subsequently by the latent cross-attention module.

\begin{figure*}[htb!]
\centering
\resizebox{\textwidth}{!}{% TikZ fragment for use with \input{axial_attention_transformer}.
% Required in the main document preamble:
% \usepackage{tikz}
% \usetikzlibrary{arrows.meta,backgrounds,calc,fit,positioning}

\definecolor{axinput}{RGB}{221,235,247}
\definecolor{axpatch}{RGB}{226,239,218}
\definecolor{axstage}{RGB}{252,228,214}
\definecolor{axmerge}{RGB}{255,242,204}
\definecolor{axlatent}{RGB}{234,223,246}
\definecolor{axattn}{RGB}{218,227,243}

\tikzset{
  axflow/.style={-{Latex[length=2.2mm]}, line width=0.75pt, draw=black!70},
  axbox/.style={draw=black!65, rounded corners=1.5pt, line width=0.75pt,
    align=center, minimum height=15mm, font=\sffamily\small},
  axsmall/.style={axbox, minimum width=26mm, minimum height=12mm,
    font=\sffamily\scriptsize},
    axln/.style={axsmall, fill=green!1, minimum width=5mm,
      minimum height=15mm, inner xsep=1pt},
  axnote/.style={font=\sffamily\scriptsize, text=black!75, align=center}
}

\begin{tikzpicture}[node distance=5mm and 6mm]
  % Panel (a): complete network.
  \node[font=\sffamily\bfseries] (titlea) {(a) Overall architecture};
  \node[axbox, fill=axinput, minimum width=25mm, right=7mm of titlea]
    (input) {HHT input\\$3M\times T\times D_{\rm{det}}$};
  \node[axbox, fill=axpatch, minimum width=29mm, right=of input]
    (patch) {Temporal patches\\$L=256$, $S=128$\\Linear $256\to180$};
  \node[axbox, fill=axpatch, minimum width=31mm, right=of patch]
    (group) {Group components\\IMF, IF, IA\\add embeddings};
  \node[axbox, fill=axstage, minimum width=32mm, right=of group]
    (component) {Component-specific\\3 independent encoders\\axial block $\times1$, $C=180$};

  \node[axbox, fill=axstage, minimum width=27mm, below=14mm of component]
    (stage1) {Stage 1\\Axial block $\times2$\\$C=180$};
  \node[axbox, fill=axmerge, minimum width=27mm, left=of stage1]
    (merge1) {Time merge\\concat pairs\\Linear $360\to180$};
  \node[axbox, fill=axstage, minimum width=27mm, left=of merge1]
    (stage2) {Stage 2\\Axial block $\times2$\\$C=180$};
  \node[axbox, fill=axmerge, minimum width=27mm, left=of stage2]
    (merge2) {Time merge\\concat pairs\\Linear $360\to320$};
  \node[axbox, fill=axstage, minimum width=27mm, left=of merge2]
    (stage3) {Stage 3\\Axial block $\times6$\\$C=320$};

  \node[axbox, fill=axmerge, minimum width=27mm, below=14mm of stage3]
    (merge3) {Time merge\\concat pairs\\Linear $640\to520$};
  \node[axbox, fill=axstage, minimum width=27mm, right=of merge3]
    (stage4) {Stage 4\\Axial block $\times2$\\$C=520$};
  \node[axln, right=4mm of stage4] (projectln) {LN};
  \node[axbox, fill=axlatent, minimum width=29mm, right=2mm of projectln]
    (project) {Projection\\Linear $520\to600$\\flatten tokens};
  \node[axbox, fill=axlatent, minimum width=31mm, right=of project]
    (mixer) {Latent mixer\\cross-attention\\32 latents, $C=600$};
  \node[axbox, fill=axlatent, minimum width=29mm, right=of mixer]
    (latentblocks) {Latent self-attention\\block $\times2$\\4 heads};

  \node[axbox, fill=axinput, minimum width=28mm, below=14mm of latentblocks]
    (head) {Attention pooling\\Linear $1024\to N_{\mathrm{class}}$\\class logits};

  \draw[axflow] (input) -- (patch);
  \draw[axflow] (patch) -- (group);
  \draw[axflow] (group) -- (component);
  \draw[axflow] (component.south) -- (stage1.north);
  \draw[axflow] (stage1) -- (merge1);
  \draw[axflow] (merge1) -- (stage2);
  \draw[axflow] (stage2) -- (merge2);
  \draw[axflow] (merge2) -- (stage3);
  \draw[axflow] (stage3.south) -- (merge3.north);
  \draw[axflow] (merge3) -- (stage4);
  \draw[axflow] (stage4) -- (projectln);
  \draw[axflow] (projectln) -- (project);
  \draw[axflow] (project) -- (mixer);
  \draw[axflow] (mixer) -- (latentblocks);
  \draw[axflow] (latentblocks.south) -- (head.north);

  \node[axnote, below=2mm of patch]
    {$N=1+\lfloor(T-L)/S\rfloor$};
  \node[axnote, below=2mm of stage1] {$N$ patches};
  \node[axnote, below=2mm of stage2] {$\lceil N/2\rceil$ patches};
  \node[axnote, below=2mm of stage3] {$\lceil N/4\rceil$ patches};
  \node[axnote, below=2mm of stage4] {$\lceil N/8\rceil$ patches};

  % Offset cards indicate repeated blocks.
  \begin{scope}[on background layer]
    \node[axbox, fill=axstage!82, minimum width=27mm]
      at ($(stage1.center)+(1.5mm,1.5mm)$) {};
    \node[axbox, fill=axstage!82, minimum width=27mm]
      at ($(stage2.center)+(1.5mm,1.5mm)$) {};
    \node[axbox, fill=axstage!72, minimum width=27mm]
      at ($(stage3.center)+(3mm,3mm)$) {};
    \node[axbox, fill=axstage!82, minimum width=27mm]
      at ($(stage3.center)+(1.5mm,1.5mm)$) {};
    \node[axbox, fill=axstage!82, minimum width=27mm]
      at ($(stage4.center)+(1.5mm,1.5mm)$) {};
    \node[axbox, fill=axlatent!82, minimum width=29mm]
      at ($(latentblocks.center)+(1.5mm,1.5mm)$) {};
  \end{scope}

  % Panel (b): one ModeTimeAttentionBlock.
  \coordinate (panelb) at ($(merge3.south west |- head.south)+(0,-24mm)$);
  \node[font=\sffamily\bfseries, anchor=west] (titleb) at (panelb)
    {(b) Axial attention block};
\tikzset{
    axop/.style={axsmall, fill=axattn, minimum width=17mm,
      minimum height=15mm, inner xsep=2pt},
    axadd/.style={circle, draw=black!65, fill=white, minimum size=4mm,
      inner sep=0pt, font=\small}
  }
  \node[axsmall, fill=axinput, minimum width=14mm, anchor=north west] (xin)
    at ($(titleb.south west)+(0,-12mm)$) {Input\\$\mathbf{x}$};
  \node[axln, right=4mm of xin] (modeln) {LN};
  \node[axop, right=2mm of modeln] (mode) {Mode\\MHA};
  \node[axadd, right=2mm of mode] (modeadd) {$+$};
  \coordinate (modeskip) at ($(xin.east)!0.5!(modeln.west)$);
  \draw[axflow] (xin.east) -- (modeln.west);
  \draw[axflow] (modeln.east) -- (mode.west);
  \draw[axflow] (mode.east) -- (modeadd.west);
  \fill[black!70] (modeskip) circle (0.6pt);
  \draw[axflow] (modeskip) -- ++(0,11mm) -| (modeadd.north);
  \node[axln, right=4mm of modeadd] (timeln) {LN};
  \node[axop, right=2mm of timeln] (time) {Time\\MHA};
  \node[axadd, right=2mm of time] (timeadd) {$+$};
  \coordinate (timeskip) at ($(modeadd.east)!0.5!(timeln.west)$);
  \draw[axflow] (modeadd.east) -- (timeln.west);
  \draw[axflow] (timeln.east) -- (time.west);
  \draw[axflow] (time.east) -- (timeadd.west);
  \fill[black!70] (timeskip) circle (0.6pt);
  \draw[axflow] (timeskip) -- ++(0,11mm) -| (timeadd.north);
  \node[axln, right=4mm of timeadd] (detectorln) {LN};
  \node[axop, right=2mm of detectorln] (detector) {Detector\\MHA};
  \node[axadd, right=2mm of detector] (detectoradd) {$+$};
  \coordinate (detectorskip) at ($(timeadd.east)!0.5!(detectorln.west)$);
  \draw[axflow] (timeadd.east) -- (detectorln.west);
  \draw[axflow] (detectorln.east) -- (detector.west);
  \draw[axflow] (detector.east) -- (detectoradd.west);
  \fill[black!70] (detectorskip) circle (0.6pt);
  \draw[axflow] (detectorskip) -- ++(0,11mm) -| (detectoradd.north);
  \node[axln, right=4mm of detectoradd] (ffnln) {LN};
  \node[axop, right=2mm of ffnln] (ffn) {FFN};
  \node[axadd, right=2mm of ffn] (ffnadd) {$+$};
  \coordinate (ffnskip) at ($(detectoradd.east)!0.5!(ffnln.west)$);
  \draw[axflow] (detectoradd.east) -- (ffnln.west);
  \draw[axflow] (ffnln.east) -- (ffn.west);
  \draw[axflow] (ffn.east) -- (ffnadd.west);
  \fill[black!70] (ffnskip) circle (0.6pt);
  \draw[axflow] (ffnskip) -- ++(0,11mm) -| (ffnadd.north);
  \node[axsmall, fill=axinput, minimum width=14mm, right=4mm of ffnadd] (xout)
    {Output\\$\mathbf{y}$};
  \draw[axflow] (ffnadd) -- (xout);
  \node[axnote, below=4mm of mode]
    {MHA:multi-head attention with 4 heads\\
    FFN:Linear-GELU-Dropout-Linear-Dropout
    };
\end{tikzpicture}}
\caption{The Transformer architecture for HHT inputs. The model contains $45{,}158{,}974$ trainable parameters for $N_{\mathrm{class}}=51$. Here, $M$, $T$, and $D_{\rm{det}}$ denote the number of HHT components (IMFs, IFs, and IAs), time length, and number of detectors, respectively. $L$, $S$, and $C$ denote the temporal patch length, stride, and feature dimension, respectively. For each temporal window and detector, the IMF, IF, and IA are independently projected into token embeddings by a linear layer. The hierarchical axial attention blocks share their parameters across the IMF, IF, and IA components. Each component is nevertheless processed independently within these blocks; the axial self-attention operations mix information only along the mode, temporal-patch, and detector axes. Information across the three components is integrated subsequently by the latent cross-attention module.}
\label{fig:axial_attention_transformer}
\end{figure*}
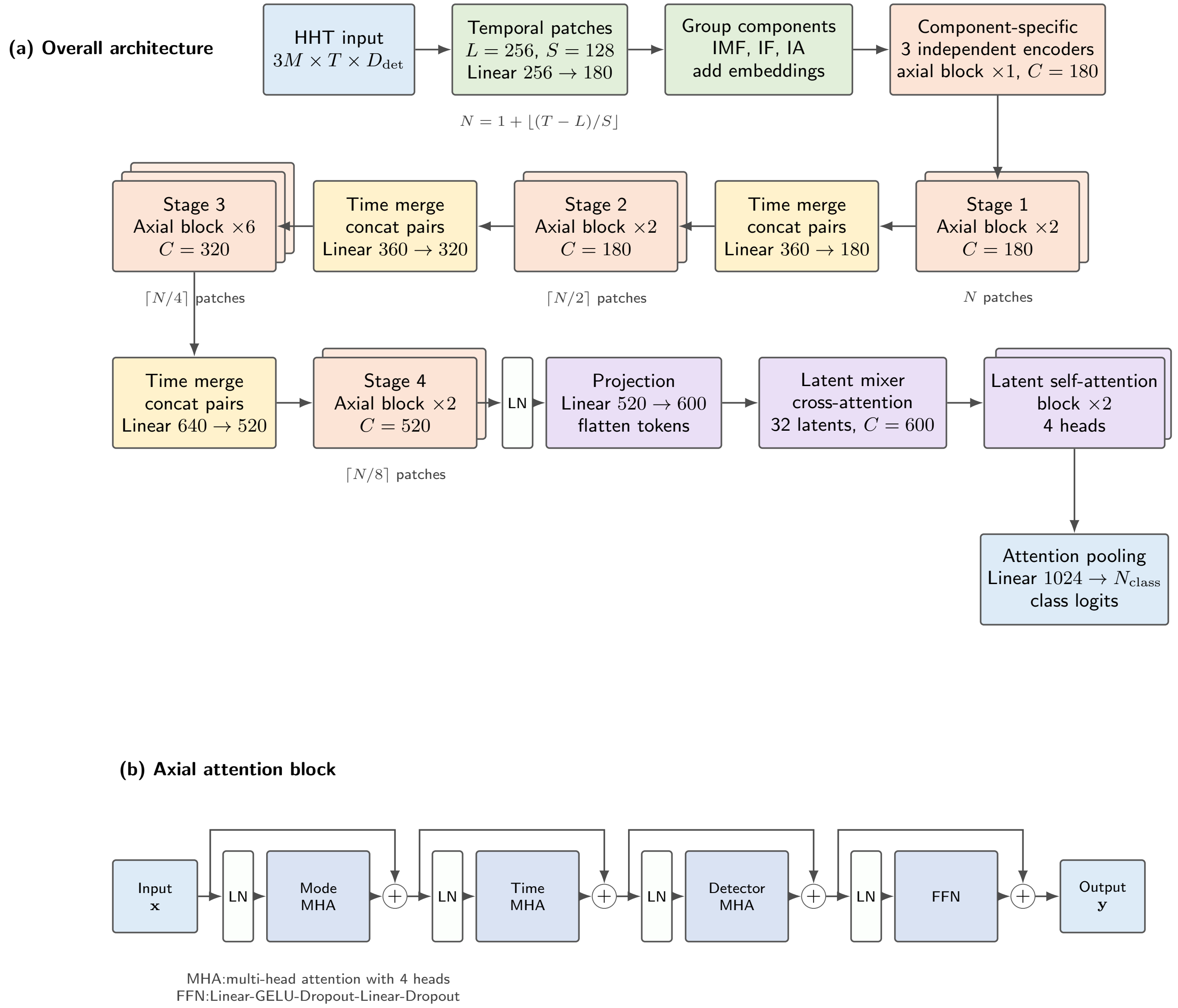

\begin{acknowledgments}
This research was supported in part by the Japan Society for the Promotion of Science (JSPS) Postdoctoral Fellowships for Research in Japan No.~P25701 (C.~All\'en\'e), the JSPS Grant-in-Aid for Fellows No.~P21726 (M.~Meyer-Conde), the JSPS Grant-in-Aid for Scientific Research No.~25K07329 (K.~Sakai) and the JSPS Grant-in-Aid for Scientific Research Nos.~23H01176, 23K25872, and 23K22499 (H.~Takahashi). This research was supported by the Joint Research Program of the Institute for Cosmic Ray Research, The University of Tokyo, and the Tokyo City University Prioritized Studies and Research Equipment Program.

This research has made use of data or software obtained from a service of the LIGO Scientific Collaboration, the Virgo Collaboration, and KAGRA.
\end{acknowledgments}

\bibliographystyle{apsrev4-2}
\bibliography{main}

\end{document}